\documentstyle[psfig]{mn}

\newif\ifAMStwofonts

\def\ApJ{ApJ }
\def\AAA{A\&A }

\def\ARAA{ARA\&A }

\def\MNRAS{MNRAS }

\def\hc5n{HC$_5$N }
\def\h2s{H$_2$S}
\def\so2{SO$_2$}

\ifoldfss
  \ifCUPmtlplainloaded \else
    \NewTextAlphabet{textbfit} {cmbxti10} {}
    \NewTextAlphabet{textbfss} {cmssbx10} {}
    \NewMathAlphabet{mathbfit} {cmbxti10} {} 
    \NewMathAlphabet{mathbfss} {cmssbx10} {} 
  \fi
  \ifAMStwofonts
    \ifCUPmtlplainloaded \else
      \NewSymbolFont{upmath} {eurm10}
      \NewSymbolFont{AMSa} {msam10}
      \NewMathSymbol{\upi}     {0}{upmath}{19}
      \NewMathSymbol{\umu}     {0}{upmath}{16}
      \NewMathSymbol{\upartial}{0}{upmath}{40}
      \NewMathSymbol{\leqslant}{3}{AMSa}{36}
      \NewMathSymbol{\geqslant}{3}{AMSa}{3E}

       \let\le=\leqslant
       \let\ge=\geqslant
    \fi
  \fi
\fi 

\ifnfssone
  \newmathalphabet{\mathit}
  \addtoversion{normal}{\mathit}{cmr}{m}{it}
  \addtoversion{bold}{\mathit}{cmr}{bx}{it}
  \newmathalphabet{\mathbfit} 
  \addtoversion{normal}{\mathbfit}{cmr}{bx}{it}
  \addtoversion{bold}{\mathbfit}{cmr}{bx}{it}
  \newmathalphabet{\mathbfss} 
  \addtoversion{normal}{\mathbfss}{cmss}{bx}{n}
  \addtoversion{bold}{\mathbfss}{cmss}{bx}{n}
  \ifAMStwofonts
    \ifCUPmtlplainloaded \else
      \UseAMStwoboldmath
      \makeatletter
      \new@mathgroup\upmath@group
      \define@mathgroup\mv@normal\upmath@group{eur}{m}{n}
      \define@mathgroup\mv@bold\upmath@group{eur}{b}{n}
      \edef\UPM{\hexnumber\upmath@group}
      \new@mathgroup\amsa@group
      \define@mathgroup\mv@normal\amsa@group{msa}{m}{n}
      \define@mathgroup\mv@bold\amsa@group{msa}{m}{n}
     \edef\AMSa{\hexnumber\amsa@group}
      \makeatother
      \mathchardef\upi="0\UPM19
      \mathchardef\umu="0\UPM16
      \mathchardef\upartial="0\UPM40
      \mathchardef\leqslant="3\AMSa36
      \mathchardef\geqslant="3\AMSa3E

       \let\le=\leqslant
       \let\ge=\geqslant
    \fi
  \fi
\fi 

\ifnfsstwo
  \DeclareMathAlphabet{\mathbfit}{OT1}{cmr}{bx}{it}
  \SetMathAlphabet\mathbfit{bold}{OT1}{cmr}{bx}{it}
  \DeclareMathAlphabet{\mathbfss}{OT1}{cmss}{bx}{n}
  \SetMathAlphabet\mathbfss{bold}{OT1}{cmss}{bx}{n}
  \ifAMStwofonts
    \ifCUPmtlplainloaded \else
      \DeclareSymbolFont{UPM}{U}{eur}{m}{n}
      \SetSymbolFont{UPM}{bold}{U}{eur}{b}{n}
      \DeclareSymbolFont{AMSa}{U}{msa}{m}{n}
      \DeclareMathSymbol{\upi}{0}{UPM}{"19}
      \DeclareMathSymbol{\umu}{0}{UPM}{"16}
      \DeclareMathSymbol{\upartial}{0}{UPM}{"40}
      \DeclareMathSymbol{\leqslant}{3}{AMSa}{"36}
      \DeclareMathSymbol{\geqslant}{3}{AMSa}{"3E}

       \let\le=\leqslant
       \let\ge=\geqslant
    \fi
  \fi
\fi 

\ifCUPmtlplainloaded \else
  \ifAMStwofonts \else 
    \def\upi{\pi}
    \def\umu{\mu}
    \def\upartial{\partial}
  \fi
\fi

\title{Chemical differentiation along the CepA-East outflows} 
\author[C. Codella et al.]
       {C. Codella,$^1$ R. Bachiller,$^2$ M. Benedettini,$^{3,4}$ P. Caselli,$^5$ S. Viti$^4$ and V. Wakelam$^{6,7}$\\
$^1$Istituto di Radioastronomia, INAF, Sezione di Firenze, Largo E. Fermi 5,
50125 Firenze, Italy\\
$^2$Observatorio Astron\'omico Nacional (IGN), Apartado 1143,
E-28800, Alcal\'a de Henares (Madrid), Spain\\
$^3$Istituto di Fisica dello Spazio Interplanetario, INAF,
Area di Ricerca Tor Vergata, Via Fosso del Cavaliere 100, 00133 Roma, Italy\\
$^4$Department of Physics and Astronomy, University College London, Gower Street WC1E6 BT London, UK\\
$^5$Osservatorio Astrofisico di Arcetri, INAF, Largo E. Fermi 5, 50125 Firenze, Italy\\
$^6$Observatoire de Bordeaux, BP 89, 33270 Floirac, France\\
$^7$The Ohio State University, Department of Physics, 174 W. 18th Ave., Columbus, OH 43210-1106 USA}
\date{Accepted date. 
      Received date;
      in original form date}

\pagerange{\pageref{firstpage}--\pageref{lastpage}}
\pubyear{date}

\begin{document}

\maketitle

\label{firstpage}

\begin{abstract}

We present the results of a multiline survey at mm-wavelengths of the Cepheus A
star forming region. 
Four main flows have been identified:
three pointing in the SW, NE, and SE directions and accelerating high density CS clumps. 
The fourth outflow, revealed by high-sensitivity HDO observations, 
is pointing towards South and
is associated with conditions particularly favourable to a chemical enrichment.
At the CepA-East position the emissions due to the ambient clump and to the
outflows coexist and   
different molecules exhibit different spectral behaviours.
Some species (C$^{13}$CH, C$_3$H$_2$, CH$_2$CO, CH$_3$C$_2$H, HC$^{18}$O$^+$)
exhibit relatively narrow lines at ambient velocities (ambient peak).
Other molecules (CO, CS, H$_2$S, SiO, SO, SO$_2$)
show extended wings tracing the whole range of the outflow velocities. 
Finally, OCS, H$_2$CS, HDO, and CH$_3$OH are associated with
wings and, in addition, show wings and in addition reveal a bright high
velocity redshifted spectral peak (outflow peak)  
which can be used to investigate the southern outflows.
At ambient velocities the gas is dense ($>$ 10$^5$ cm$^{-3}$)
and different components at distinct temperatures coexist, ranging from the
relatively low kinetic temperatures ($\le$ 50 K)
measured with H$_2$S, CH$_3$OH, H$_2$CS, and CH$_3$C$_2$H, to
definitely higher temperature conditions, $\sim$ 100-200 K,
obtained from the SiO, SO, and SO$_2$ spectra. 
For the outflow peak we derive densities between $\sim$ 10$^4$ cm$^{-3}$ to
$\sim$ 10$^7$ cm$^{-3}$ and high temperatures, $\simeq$ 100-200 K, 
indicating regions compressed and heated by shocks.

The analysis of the line profiles shows that 
the SiO molecule dominates at the highest velocities and 
at the highest excitation conditions, confirming its
close association with shocks.
H$_2$S, SO$_2$, and SO preferentially 
trace more quiescent regions than SiO, and in particular 
a lack of bright H$_2$S emission at the highest velocities is found.
OCS and H$_2$CS emit at quite high velocities, where 
the abundances of three shock tracers like SiO, CH$_3$OH, and HDO are higher. 
These results may indicate that H$_2$S is not the only major sulphur carrier in the grain mantles,
and that OCS and H$_2$CS may probably play an important role on the grains; or that 
alternatively they rapidly form once the mantle is evaporated after the passage of a shock.
Finally, the outflow peak emission has been compared with recent 
time-dependent sulphur chemistry 
models: the results indicate that, if associated with accurate measurements of the
physical conditions, the CH$_3$OH/H$_2$CS
column density ratio can be used as an effective chemical clock to date the age of shocked gas.

\end{abstract}

\begin{keywords}
ISM: clouds -- ISM: individual objects: CepA -- ISM: jets and outflows --
ISM: molecules -- Radio lines: ISM
\end{keywords}

\section[]{Introduction}

The mass loss from Young Stellar Objects (YSOs) 
produces high velocity flows which strike the
ambient medium driving shocks. Once the temperature has increased,
at least to a few thousands degrees depending on the shock type, 
the energy barriers between neutral molecules can be overcome and the chemistry
of certain species, such as the
sulphuretted species, is altered significantly 
(see e.g. Pineau des For$\hat {\rm e}$ts et al. 1993, van Dishoeck \& Blake 
1998, and reference therein). 
In addition, grains are affected by shocks, with the consequent injection
of molecular and atomic species in the gas phase: again, this leads
to an enhancement of the abundances of several species, including  
S-bearing molecules.
The scenario proposed by most models is that H$_2$S is the main reservoir of
sulphur on grain mantles: once in the gas phase,
H$_2$S is used for a fast production of SO and SO$_2$ (e.g. Pineau
des For$\hat {\rm e}$ts et al. 1993, Charnley 1997). 
However, the lack of H$_2$S features in the ISO spectra (Gibb et al. 2000,
Boogert et al. 2000) which set upper limits on the iced H$_2$S
abundance around protostars and the detection of OCS on grains (Palumbo
et al. 1997) suggest that the latter may be an  
important sulphur carrier in the ices.
This seems supported by the observations in the envelopes of 
massive young stars recently 
performed by van der Tak et al. \shortcite{vander},
which indicate for OCS higher excitation temperatures 
than for H$_2$S.
An alternative hypothesis is that the sulphur released from
the dust mantles is mainly in atomic form (Wakelam et al. 2004).
In any case, once the gas phase has been enriched by the
passage of a shock, other 
S-bearing species such as 
H$_2$CS and HCS$^+$ are expected to significantly increase their
abundances as a consequence of the sulphur injection.
Therefore, estimates of abundance ratios such as SO$_2$/H$_2$S,
H$_2$S/OCS, and H$_2$CS/OCS may provide us with  
chemical clocks to study the evolutionary stages of molecular outflows.
In fact, the use of SO$_2$/H$_2$S and H$_2$S/OCS ratios has
already led crude age estimates of the outflows located
in CB3 (Codella \& Bachiller 1999) and L1157 (Bachiller et al. 2001), encouraging further
studies.

The high temperature scenario can be applied also to hot cores around
protostars, which are characterised by high temperatures ($\ge$ 100 K),
high densities ($\ge$ 10$^6$ cm$^{-3}$), and a rich molecular inventory.
The SO$_2$/H$_2$S ratio has been used by Hatchell et al. \shortcite{hatchell} 
for massive hot cores: the inferred ages are in agreement 
with the dynamical times estimated from the associated outflows.
Moreover, Buckle \& Fuller \shortcite{buckle} show that the chemical
evolution of sulphuretted species is a potential probe of timescales also in
low-mass star forming regions.
In summary, the effect of high-temperature chemistry on 
the composition of the gas hosting the star forming process
can be used as a tool to investigate the evolution of protostars. 

\begin{table*}
\caption[] {List of molecular species, transitions and observing parameters}
\begin{tabular}{lrrcccccc}
\hline
\multicolumn{1}{c}{Transition} &
\multicolumn{1}{c}{$\nu_{\rm 0}$} &
\multicolumn{1}{c}{$E_{\rm u}$} &
\multicolumn{1}{c}{$A_{\rm ul}$} &
\multicolumn{1}{c}{$F_{\rm int}$} &
\multicolumn{1}{c}{$HPBW$} &
\multicolumn{1}{c}{$T_{\rm sys}$} &
\multicolumn{1}{c}{$dv ({\rm AC})$} &
\multicolumn{1}{c}{$dv ({\rm 1 MHz})$} \\
\multicolumn{1}{c}{} &
\multicolumn{1}{c}{(MHz)} &
\multicolumn{1}{c}{(K)} &
\multicolumn{1}{c}{(s$^{-1}$)} &
\multicolumn{1}{c}{(K km/s)} &
\multicolumn{1}{c}{($\arcsec$)} &
\multicolumn{1}{c}{(K)} &
\multicolumn{1}{c}{(km/s)} &
\multicolumn{1}{c}{(km/s)} \\
\hline
\multicolumn{9}{c}{Selected transitions} \\
\hline
HDO($J_{\rm K_-K_+}$ = 1$_{\rm 10}$--1$_{\rm 11}$) & 80578.30 & 47 & 1.3 10$^{-6}$ & 0.58(0.05) & 31 & 150 & 0.15 & 3.72 \\
OCS($J$ = 7--6) & 85139.12 & 16 & 1.7 10$^{-6}$ & 0.36(0.02) & 29 & 110 & 0.14 & -- \\ 
HCS$^+$($J$ = 2--1) & 85347.88 & 6 & 1.1 10$^{-5}$ & 0.38(0.04) & 29 & 110 & 0.14 & 3.51 \\ 
CH$_3$C$_2$H($J_{\rm K}$ = 5$_{\rm K}$--4$_{\rm K}$) & 85457.30$^a$ & 12$^a$ & 2.0 10$^{-6}$$^a$ & 3.11(0.13)$^b$ & 29 & 150 & 0.27 & 3.51 \\
C$^{34}$S($J$ = 2--1) & 96412.98 & 7 & 1.6 10$^{-5}$ & 2.16(0.18) & 26 & 220 & 0.24 & 3.11 \\
CS($J$ = 2--1) & 97980.97 & 7 & 1.7 10$^{-5}$ & 28.03(0.14) & 25 & 180 & 0.24 & 3.06 \\
SO($J_{\rm K}$ = 3$_2$--2$_1$) & 99299.88 & 4 & 1.1 10$^{-5}$ & 14.42(0.03) & 24 & 150 & 0.24 & 3.02 \\
H$_2$CS($J_{\rm K_-K_+}$ = 3$_{13}$--2$_{12}$) & 101477.75 & 23 & 1.3 10$^{-5}$ & 1.01(0.03) & 24 & 187 & 0.12 & 2.95 \\
$^{34}$SO$_2$($J_{\rm K_-K_+}$ = 3$_{13}$--2$_{02}$) & 102031.91 & 8 & 9.5 10$^{-6}$ & 0.11(0.01) & 24 & 130 & 0.11 & 2.94 \\ 
CH$_3$C$_2$H($J_{\rm K}$ = 6$_{\rm K}$--5$_{\rm K}$) & 102547.98$^a$ & 17$^a$ & 3.4 10$^{-6}$$^a$ & 4.16(0.14)$^b$ & 24 & 200 & 0.23 & 2.90 \\
SO($J_{\rm K}$ = 4$_3$--3$_2$) & 138178.64 & 9 & 3.1 10$^{-5}$ & 22.76(0.19) & 17 & 370 & 0.17 & 2.17 \\
CH$_3$OH($J_{\rm K}$ = 3$_{\rm K}$--2$_{\rm K}$) & 145103.23$^a$ & 14$^a$ & 1.2 10$^{-5}$$^a$ & 5.64(0.15)$^a$ & 17 & 280 & 0.08 & 2.07 \\ 
H$_2$$^{34}$S($J_{\rm K_-K_+}$ = 1$_{10}$--1$_{01}$) & 167910.52 & 28 & 2.6 10$^{-5}$ & 1.11(0.23) & 14 & 480 & 0.07 & 1.79 \\ 
H$_2$CS($J_{\rm K_-K_+}$ = 6$_{16}$--5$_{15}$) & 202923.55 & 47 & 1.2 10$^{-4}$ & 2.15(0.08) & 12 & 460 & 0.06 & 1.48 \\
CH$_3$C$_2$H($J_{\rm K}$ = 12$_{\rm K}$--11$_{\rm K}$) & 205080.73$^a$ & 64$^a$ & 2.8 10$^{-5}$$^a$ & 6.71(0.36)$^b$ & 12 & 600 & 0.11 & 1.46 \\
HCS$^+$($J$ = 5--4) & 213360.64 & 31 & 2.0 10$^{-4}$ & 0.59(0.08) & 12 & 330 & 0.04 & 1.41 \\
H$_2$$^{34}$S($J_{\rm K_-K_+}$ = 2$_{20}$--2$_{11}$) & 213376.92 & 84 & 1.7 10$^{-5}$ & $\le$0.07$^c$ & 12 & 330 & 0.04 & 1.41 \\
SiO($J$ = 5--4) & 217104.94 & 31 & 5.2 10$^{-4}$ & 8.71(0.24) & 11 & 450 & 0.43 & 1.38 \\
C$^{18}$O($J$ = 2--1) & 219560.33 & 16 & 6.2 10$^{-7}$ & 47.94(0.25) & 11 & 580 & 0.43 & 1.36 \\
SO($J_{\rm K}$ = 6$_5$--5$_4$) & 219949.39 & 24 & 1.3 10$^{-4}$ & 40.74(0.30) & 11 & 580 & 0.11 & 1.36 \\
CO($J$ = 2--1) & 230537.98 & 17 & 6.9 10$^{-7}$ & --$^d$ & 10 & 1100 & 0.10 & 1.30 \\ 
C$^{34}$S($J$ = 5--4) & 241016.17 & 34 & 2.8 10$^{-4}$ & 1.75(0.40) & 10 & 990 & 0.10 & 1.24 \\
HDO($J_{\rm K_-K_+}$ = 2$_{\rm 11}$--2$_{\rm 12}$) & 241561.53 & 95 & 1.2 10$^{-5}$ & 3.42(0.19) & 10 & 68 & 0.05 & 1.24 \\
CH$_3$OH($J_{\rm K}$ = 5$_{\rm K}$--4$_{\rm K}$) & 241791.44$^a$ & 35$^a$ & 5.8 10$^{-5}$$^a$ & 8.32(0.07)$^a$ & 10 & 678 & 0.05 & 1.24 \\ 
H$_2$CS($J_{\rm K_-K_+}$ = 7$_{16}$--6$_{15}$) & 244047.75 & 60 & 2.1 10$^{-4}$ & 1.36(0.06) & 10 & 890 & 0.05 & 1.23 \\
CS($J$ = 5--4) & 244935.61 & 34 & 3.0 10$^{-4}$ & 47.33(0.53) & 10 & 1100 & 0.10 & 1.22 \\ 
SO($J_{\rm K}$ = 7$_6$--6$_5$) & 261843.72 & 35 & 2.2 10$^{-4}$ & 52.30(0.58) & 9 & 1300 & 0.09 & 1.15 \\
\hline
\multicolumn{9}{c}{Serendipity detections} \\
\hline
HC$^{18}$O$^+$($J$ = 1--0) & 85162.21 & 4 & 3.6 10$^{-5}$ & 0.65(0.02) & 29 & 110 & 0.14 & -- \\ 
C$^{13}$CH($J_{\rm K_-K_+}$ = 1$_{11}$--0$_{11}$) & 85307.69 & 4 & 1.3 10$^{-6}$ & 0.18(0.03) & 29 & 110 & -- & 3.51 \\
C$_3$H$_2$($J_{\rm K_-K_+}$ = 2$_{12}$--1$_{01}$) & 85338.91 & 6 & 2.3 10$^{-5}$ & 1.18(0.05) & 29 & 110 & 0.14 & 3.51 \\
CH$_2$CO($J_{\rm K_-K_+}$ = 5$_{14}$--4$_{13}$) & 101981.43 & 23 & 1.1 10$^{-5}$ & 0.14(0.01) & 24 & 130 & 0.11 & 2.94 \\ 
CH$_3$OH($J_{\rm K}$ = 10$_{\rm -2}$--10$_{\rm 1}$ E) & 102122.70 & 154 & 1.7 10$^{-7}$ & 0.08(0.01) & 24 & 130 & -- & 2.94 \\ 
CH$_3$OH($J_{\rm K}$ = 9$_{\rm 1}$--9$_{\rm 0}$ E) & 167931.13 & 126 & 2.3 10$^{-5}$ & 2.33(0.25) & 14 & 480 & 0.07 & 1.79 \\ 
CH$_3$OH($J_{\rm K}$ = 13$_{\rm 6}$--14$_{\rm 5}$ E) & 213377.52 & 390 & 1.1 10$^{-5}$ & 0.43(0.03) & 12 & 330 & 0.04 & 1.41 \\
CH$_3$OH($J_{\rm K}$ = 1$_{\rm 1}$--0$_{\rm 0}$ E) & 213427.12 & 13 & 3.4 10$^{-5}$ & 2.20(0.06) & 12 & 330 & 0.04 & 1.41 \\
$^{34}$SO$_2$($J_{\rm K_-K_+}$ = 16$_{115}$--15$_{214}$) & 241509.05 & 131 & 8.3 10$^{-5}$ & 0.78(0.14) & 10 & 670 & -- & 1.24 \\
HNCO($J_{\rm K_-K_+}$ = 11$_{\rm 011}$--10$_{\rm 010}$) & 241774.09 & 58 & 2.0 10$^{-4}$ & 2.81(0.20) & 10 & 678 & 0.05 & 1.24 \\ 
$^{34}$SO$_2$($J_{\rm K_-K_+}$ = 18$_{117}$--18$_{018}$) & 243935.88 & 163 & 7.0 10$^{-5}$ & 0.61(0.73) & 10 & 890 & -- & 1.23 \\
\hline
\end{tabular}
\begin{center}
$^a$ For CH$_3$OH, it refers to the $J$$_{\rm 0}$--$J$--1$_{\rm 0}$ A$^+$ line, while for 
CH$_3$C$_2$H it refers to the $K$=0 component.
$^b$ For CH$_3$C$_2$H the integrated flux refers to the $K$=0,1 pattern. 
$^c$ It refers to the 3$\sigma$ noise level; $^d$ The CO(2--1) observations have been performed in wobbler mode 
loosing the information at ambient velocities in order to investigate the high-velocity wings. \\
\end{center}
\end{table*}

Physical conditions as well as time evolution
affect the sulphur chemistry in high-temperatures environment.
For instance, recently Wakelam et al.
\shortcite{wake} modeled the sulphur chemistry around hot core-like environments
and found that the obtained abundances depend not only on age but also on
the excitation conditions of the gas.
Thus, in order to use the sulphur abundance ratios as chemical clocks, care should
be taken to first constrain the gas conditions.
In the case of molecular outflows, it
is reasonable to expect different physical and/or chemical conditions
at the different velocities and thus a study of the excitation as a
function of the velocity is needed. The existing observations of sulphuretted molecules
have been carried out to date in a unsystematic way, and,
in particular, since the wing profiles are weak,
the chemical composition of the gas at high-velocity has
been poorly investigated.
With this in mind, we carried out line observations
of S-bearing species at mm-wavelengths towards a star forming region
with well defined high-velocity components.
Cepheus A (CepA) well represents such target: in particular, the East
component harbours an OB3 stellar association (Goetz et al. 1998,
and references therein) driving multiple outflows
(e.g. Narayanan \& Walker 1996, Bergin et al. 1997).
Bergin et al. \shortcite{bergin97} have
studied the chemical properties of Cepheus A
showing that a large number of molecules can be easily detected,
but their observations only trace the quiescent gas, this is likely
due to the limitation in spatial and spectral resolutions, which
does not allow the detection of the high velocity wings.

In this paper, we report the results of a deep millimeter survey
of molecular lines in CepA.
In particular, we compare line observations of
SO, HCS$^+$, H$_2$CS, OCS, H$_2$S, and SO$_2$
with profiles due to standard tracers of shocked gas,
high excitation conditions, and of ambient emission.
The main aim is to study the variation of the abundance ratios
and the excitation conditions
of S-bearing species along the line profiles. We find that
sulphuretted molecules may be good chemical clocks
to date molecular outflows and their driving protostellar sources.

\begin{figure}
\psfig{file=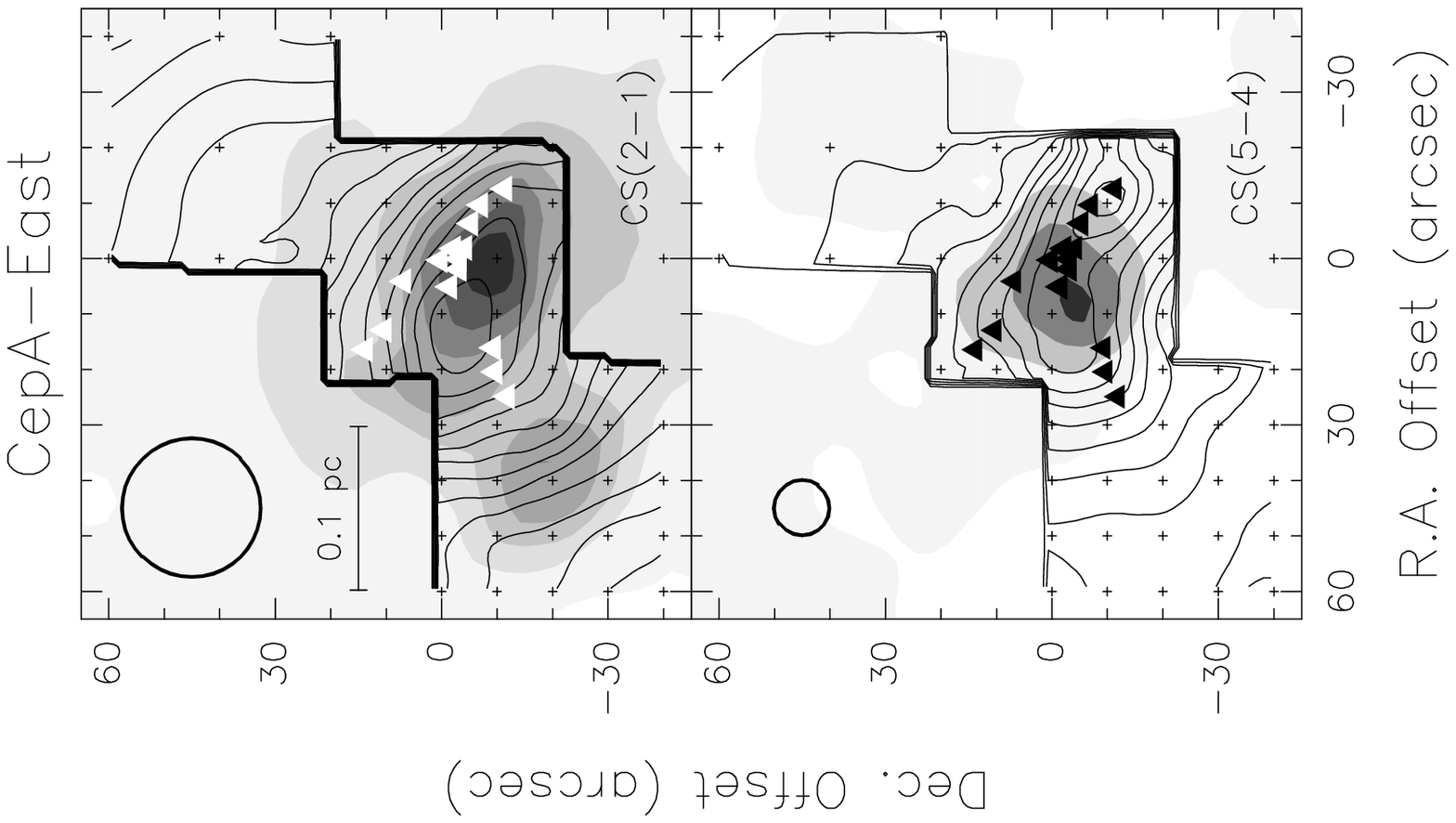,width=7cm,angle=-90}
\caption{Contour maps of the integrated CS $J$=2-1 (upper panel) and $J$=5-4 (lower
panel) integrated emission towards CepA-East.
The maps are overlaid with the grey scale images reproducing the integrated emission
of H$_2$$^{32}$S(1$_{10}$-1$_{01}$) (upper panel) and $^{32}$SO$_2$(3$_{13}$-2$_{02}$)
(lower panel), reported by Codella et al. (2003) in their Fig. 1.
The empty circles show the IRAM beam (HPBW),
while the small crosses  mark the observed positions. The triangles stand for the VLA 2 cm
continuum components which trace in the East region two strings of sources arising
in shocks (Garay et al. 1996). The velocity integration
interval is --30,+10 km s$^{-1}$. The contours range from 1.0 to 31.0 K km s$^{-1}$ (upper panel)
and from 3.0 to 57.0 K km s$^{-1}$ (lower panel). The first contours and the steps correspond
to about 3 and 6 $\sigma$, respectively (where $\sigma$ is the r.m.s. of the map).}
\end{figure}

\section[]{Observations and Results}

The observations were performed with the IRAM
30-m telescope at Pico Veleta (Granada, Spain)
in June 2001, September 2002, August 2003, and June 2004.
The observed molecular species, the transitions,
their upper level energies, the Einstein $A$ coefficients and rest frequencies and some observing parameters,
such as the HPBW and the typical system temperature ($T_{\rm sys}$)
are summarised in Table 1.
The integration time (ON+OFF source) ranged from about 1 to $\sim$18 hours,
depending on the intensities of the observed lines.
The main beam efficiency varies from about 0.8 (at 81 GHz) to 0.5 (at 262 GHz).
The observations were
made by position switching. The pointing was checked about every hour by
observing nearby planets or continuum sources and it was found to be
accurate to within 4$\arcsec$.  As spectrometers,
an autocorrelator (AC) split into different parts (up to six)
was used to allow simultaneous observations
of four different transitions.
Also a 1 MHz filter
bank, split into four parts of 256 channels, was
simultaneously used.
The velocity resolutions provided by both backends, AC and 1 MHz, are
shown in Table 1. When considered convenient, the AC spectra were smoothed to a lower velocity
resolution (up to $\sim$ 1 km s$^{-1}$).
The spectra were
calibrated with the standard chopper wheel method and are reported
here in units of main-beam brightness temperature ($T_{\rm MB}$).

The multiline molecular survey has been performed towards CepA-East and, in particular,
at the coordinates of the HW2 object, which is one of
the YSOs thought to drive the molecular outflows: 
$\alpha_{\rm 2000}$ = 22$^{\rm h}$ 56$^{\rm m}$ 17$\fs$9,
$\delta_{\rm 2000}$ = +62$\degr$ 01$\arcmin$ 49$\farcs$7.
In addition, we present maps of CepA-East in CS, a well known tracer of high density clumps,
and in HDO, which traces hot gas chemistry as well as shocked material
(e.g. van Dishoeck \& Blake 1998).
In particular, the heavy water emission will be here used as alternative to H$_2$O,
whose observations are prevented from the ground due to very strong
atmospheric absorption.
Water is expected to be one of the most important coolant
in non-dissociative shocks (e.g. Kaufman \& Neufeld 1996), since its
gas-phase abundance is considerably enhanced both via high temperature 
(larger than 200-300 K) reactions and by sputtering. 
Given the characteristics of CepA-East, it will thus be possible
to examine in details how the dense medium is affected by the 
occurrence of YSOs 

\begin{figure}
\psfig{file=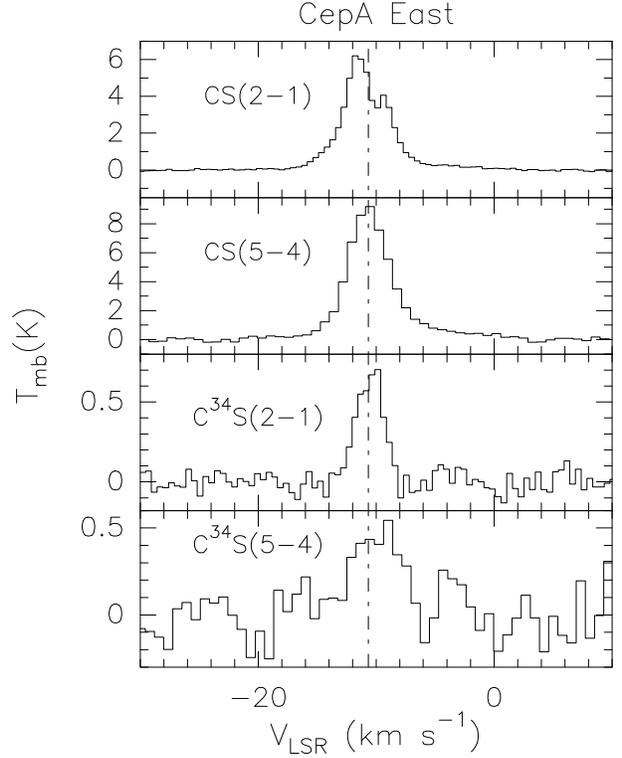,width=8cm,angle=-90}
\caption{
CS and C$^{34}$S line profiles observed towards CepA East: transitions are reported.
The dashed line stands for the ambient LSR velocity (--10.65 km s$^{-1}$), according to the
CS $J$ = 5--4 and C$^{18}$O $J$ = 2--1 measurement (see text).}
\end{figure}

\begin{figure}
\psfig{file=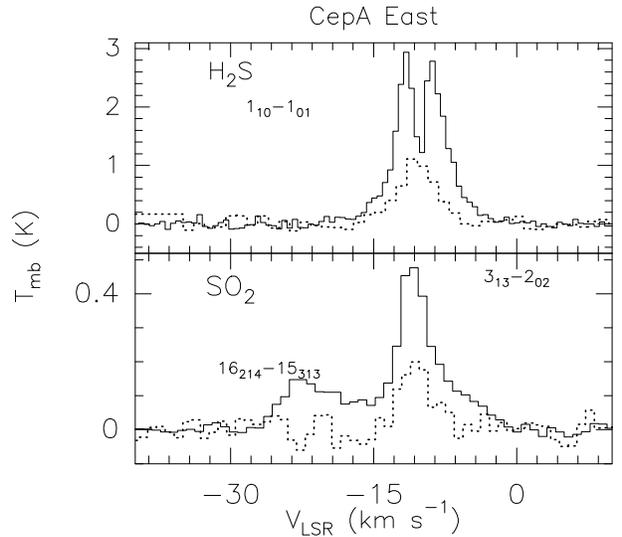,width=8cm,angle=-90}
\caption{CepA East: comparison between same line transitions
of different isotopomers.
The solid lines are for the H$_2$$^{32}$S(1$_{10}$-1$_{01}$) and
$^{32}$SO$_2$(3$_{13}$-2$_{02}$) profiles reported by Codella et al. (2003).
The dotted lines show the corresponding
H$_2$$^{34}$S and $^{34}$SO$_2$ spectra,
rescaled to be comparable with the $^{32}$S-lines. Note that only
for $^{32}$SO$_2$, the 3$_{13}$-2$_{02}$ line is blended with the
16$_{214}$-15$_{313}$ emission.}
\end{figure}

Table 1 reports the selected transitions and their flux $F_{\rm int}$ (K km s$^{-1}$)
integrated along the whole profile: all but 
H$_2$$^{34}$S(2$_{20}$--2$_{11}$) have been detected.
The maps of CepA in H$_2$S(1$_{10}$--1$_{01}$, 2$_{20}$--2$_{11}$) and 
SO$_2$(3$_{13}$--2$_{02}$, 5$_{24}$--4$_{13}$, 16$_{214}$--15$_{313}$) 
had been already reported in a previous paper (Codella et al. 2003).
The results indicate the occurrence of 
a rich chemistry associated with
CepA-East, confirming the findings of  
Bergin et al. \shortcite{bergin97}. 
The gas surrounding this region 
is particularly rich in the S-bearing molecules (SO, H$_2$S, SO$_2$, H$_2$CS, HCS$^+$, OCS). 
Also line spectra due to tracers of the high density clumps hosting the YSOs
(CS, C$^{18}$O, and CH$_3$C$_2$H) as well as to tracers of hot and/or shocked material
(SiO, HDO, CH$_3$OH) have been obtained.
Moreover, the use of large bandwidths allowed us to serendipitously detect
a number of other emission lines due to HC$^{18}$O$^+$, C$_3$H$_2$, C$^{13}$CH, CH$_2$CO,
CH$_3$OH, and HNCO, as shown in the lower part of Tab. 1.
Table 2 reports the lines which have not been identified, listing the
observed peak frequency, the spectral resolution,
the peak temperature and the FWHM linewidth. 

Table 1 shows that the lines detected are associated with a wide range
of excitation, from fews to hundreds of Kelvins. In particular, the detection of
emission due to transitions above 100 K clearly suggest the presence of high
temperature conditions, which could be associated to the presence of a hot core
and/or the occurrence of shocked material, and consequently requires the
analysis of the maps and of the line profiles.

\begin{table}
\caption[] {Unidentified detections observed towards CepA-East}
\begin{tabular}{ccrc}
\hline
\multicolumn{1}{c}{Obs. frequency} &
\multicolumn{1}{c}{$d\nu$} &
\multicolumn{1}{c}{$T_{\rm MB}$} &
\multicolumn{1}{c}{$FWHM$} \\
\multicolumn{1}{c}{(MHz)} &
\multicolumn{1}{c}{(MHz)} &
\multicolumn{1}{c}{(mK)} &
\multicolumn{1}{c}{(km/s)} \\
\hline
U--102062.23(0.15) & 0.31 & 15(3) & 7.8(2.1) \\
U--202811.00(0.19) & 1.00 & 96(20) & 5.3(0.9) \\
U--213303.65(0.14) & 1.00 & 101(12) & 8.4(1.6) \\
U--216758.40(0.47) & 1.00 & 80(19) & 9.5(1.8) \\
\hline
\end{tabular}
\end{table}

\section[]{A view of CepA-East: dense clumps and outflows}

\subsection[]{Ambient emission}

We produced relatively
small CS maps of CepA-East in order to carefully study the distribution
of the ambient clumps. In particular, we focused our attention
on the regions where the two strings of VLA continuum sources associated with
shocks (Garay et al. 1996) and the H$_2$S and SO$_2$ clumps (Codella et al. 2003) are located. 
Figure 1 shows the maps of the integrated emission due to CS $J$=2--1 (upper panel)
and $J$=5--4 (lower panel) and compare such distributions with those
of H$_2$S $J_{\rm K_-K_+}$ = 1$_{10}$--1$_{01}$ (upper panel) and
SO$_2$ $J_{\rm K_-K_+}$ = 3$_{13}$--2$_{02}$ (lower panel), reported in gray scale.
The triangles stand for the VLA objects.
It is possible to note that the two CS maps peak at different positions,
probably due to the self-absorption of the 2--1 line (see the self-reversed profile
in Fig. 2).
Also the H$_2$S and SO$_2$ emissions show a different distribution with respect to CS,
indicating that such molecules are not simply reproducing the same gas distribution and
are able to trace different gas associated with different physical
and/or chemical conditions.

Figure 2 reports the CS (upper panels) and C$^{34}$S (lower panels)
spectra due to the $J$=2--1 and 5--4 transitions.
Although the CS(5--4) line shows the occurrence of wings which modify
the gaussian profile at about 15\% of the line peak, such emission
can be used to define the characteristics of the ambient emission:
the LSR velocity is --10.65 km s$^{-1}$ (dashed line in Fig. 2),
while the FWHM is 3.8 km s$^{-1}$.
The comparison between these
CS spectra confirms what found in the recent paper
by Bottinelli \& Williams \shortcite{botti} which clearly indicates that
the CS(2--1) profile is affected by self-absorption, whereas the CS(5--4)
and the two transitions of C$^{34}$S, due to higher excitation and/or
smaller abundance, are definitely optically thinner and thus
well centered at the LSR ambient velocity.
Moreover, as reported also by Bottinelli \& Williams \shortcite{botti}, the
CS(2--1) line shows a blueshifted peak brighter than the redshifted one, suggesting
that infall motions can play an important role in the dynamics of the material
traced by CS.

\begin{figure*}
\psfig{file=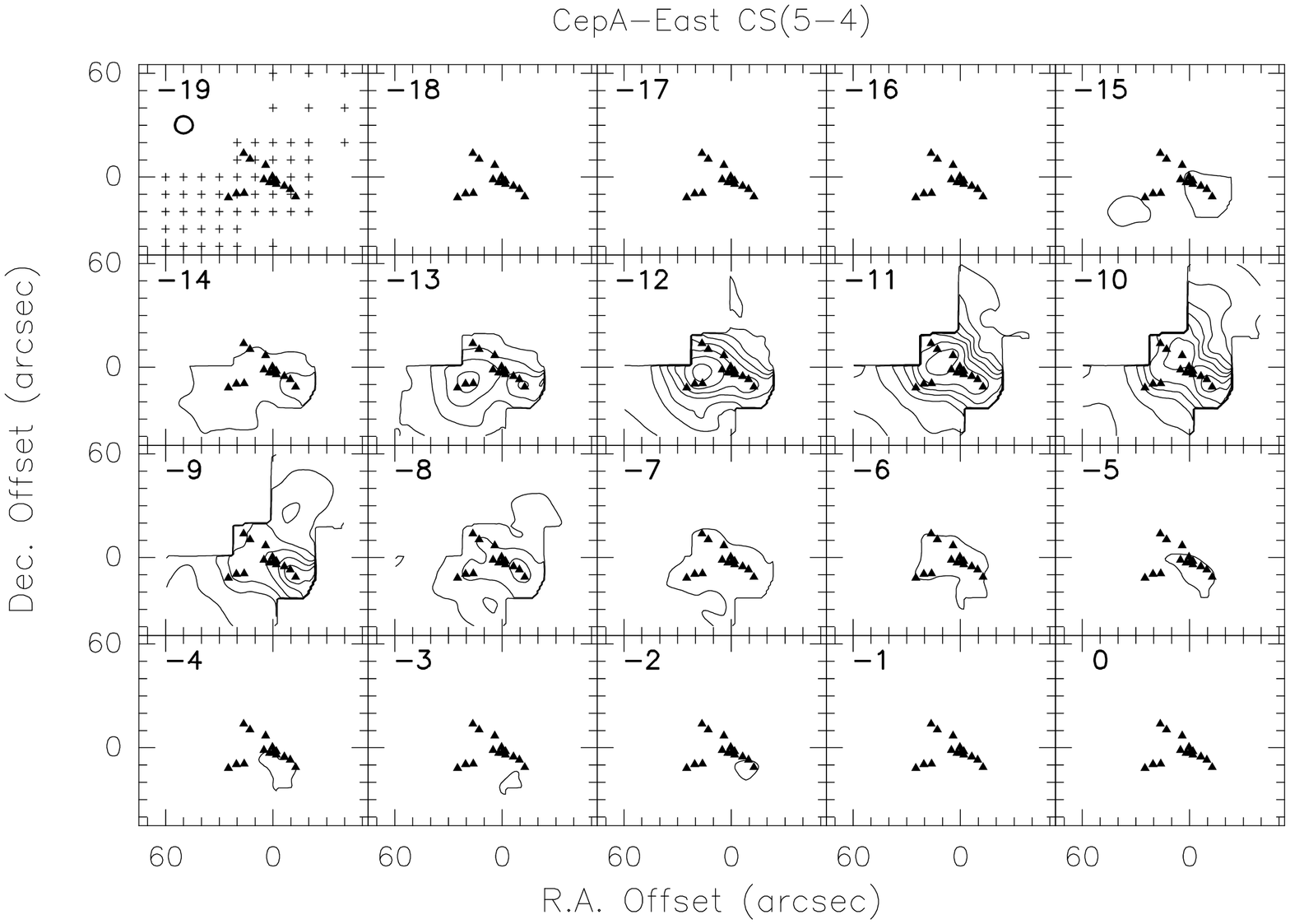,width=15cm,angle=-90}
\caption{Channel map of the
CS $J$ = 5--4 emission towards
CepA-East. Each panel shows the emission integrated over a velocity interval
of 1 km s$^{-1}$ centred at the value given in the left corner.
Symbols are drawn as in Fig. 1.
The ambient velocity emission is --10.7 km s$^{-1}$ according
to CS $J$ = 5--4 and C$^{18}$O $J$ = 2--1 measurements (see Sect. 3.2).
The contours range from 0.75 ($\sim$5$\sigma$) to 9.75
K km s$^{-1}$ by step of 1.50 K km s$^{-1}$.}
\end{figure*}

Figure 3 compares the H$_2$$^{34}$S(1$_{10}$-1$_{01}$) and the
$^{34}$SO$_2$(3$_{13}$-2$_{02}$) spectra (dotted lines, see also Fig. 1) with the profiles
of the main isotopomers (continuous lines),
reported in a previous paper (Codella et al. 2003).
Note that the SO$_2$(3$_{13}$-2$_{02}$) profile is blended with the
16$_{214}$-15$_{313}$ line.
The comparison between the H$_2$S spectra indicates that the double-peak H$_2$$^{32}$S
pattern is caused by self-absorption, with the H$_2$$^{34}$S line peaking
at the LSR velocity, where H$_2$$^{32}$S shows the relative minimum.
On the other hand, Fig. 3 confirms that the detected SO$_2$ lines do not present
self-absorption effects and they peak at the ambient velocity.

\subsection[]{Kinematics}

Figure 1 shows an elongated structure oriented in the NE-SW direction
in the CS(5--4) line, which is not affected by self-absorption and it has been
observed with an angular resolution of 10$\arcsec$.
To further study this distribution, a CS(5-4) channel map is reported in Fig. 4.
Focusing the attention on the emission at velocities close to the ambient one,
which corresponds to --10.7 with a FWHM of 1.8 km s$^{-1}$,
we note that the elongated structure is due to the presence of two clumps
centered respectively at the (+20$\arcsec$,0$\arcsec$) and (--10$\arcsec$,--10$\arcsec$)
offset map positions and with a beam deconvolved size
of about 15$\arcsec$ (0.05 pc at a distance of 725 pc, Sargent 1977)
at slightly different velocities.
This picture resembles the distribution of another high-density
tracers such as N$_2$H$^+$ (Bergin et al. 1997):
the two clumps could indicate star forming sites or could point to gas components accelerated by
the interaction with the mass loss processes indicated by the NE-SW VLA string.
The latter case seems to be more likely since the clumps are
symmetrically located with respect to the positions of the driving YSOs (the HW2 object
is located at the centre of the map) and lie along the direction
corresponding to the main axis of an extremely
high-velocity outflow ($v$--$v_{\rm LSR}$$\ge$20 km s$^{-1}$;
e.g. Narayanan \& Walker 1996).

On the other hand, from Fig. 4 it is possible to see (i) redshifted (up to --2 km s$^{-1}$)
and blueshifted (up to --15 km s$^{-1}$) emission
associated with the SW jet, and (ii) a blueshifted clump 
located at the end of the SE VLA chain
as clearly reported in the --15 km s$^{-1}$ panel. The latter clump is also
seen in H$_2$S at high velocities (see Fig. 4 of Codella et al. 
2003) and visible in Fig. 1 (upper panel) at (+40$\arcsec$,--20$\arcsec$), 
confirming that at this position we are tracing high density material 
associated with outflow motions.
In conclusion, the CS emission traces two outflow directions, the SW and SE ones, 
associated with two strings of shocked sources. 

\begin{figure}
\psfig{file=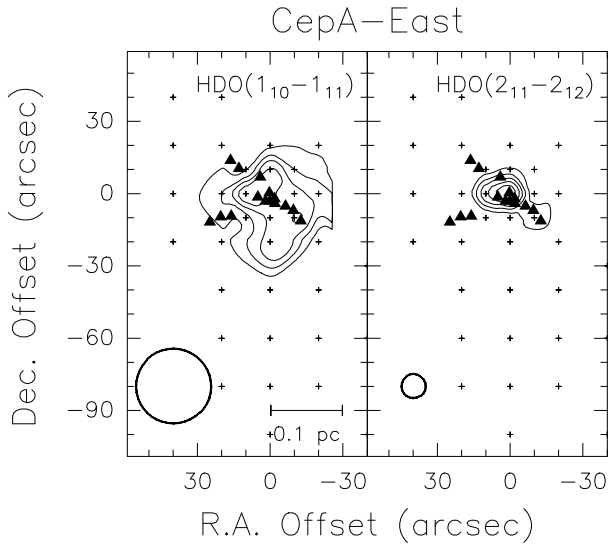,width=8cm,angle=-90}
\caption{Contour maps of the integrated HDO $J_{\rm K_-K_+}$ = 1$_{\rm 10}$--1$_{\rm 11}$
(upper panel) and $J_{\rm K_-K_+}$ = 2$_{\rm 11}$--2$_{\rm 12}$ (lower
panel) integrated emission towards CepA-East.
Symbols are drawn as in Fig. 1.
The velocity integration
interval is --20,0 km s$^{-1}$. The contours range from 0.14 to 0.56 K km s$^{-1}$
(upper panel)
and from 0.60 to 3.00 K km s$^{-1}$ (lower panel). The first contours and the steps
correspond to about 3 $\sigma$.}
\end{figure}

\begin{figure*}
\psfig{file=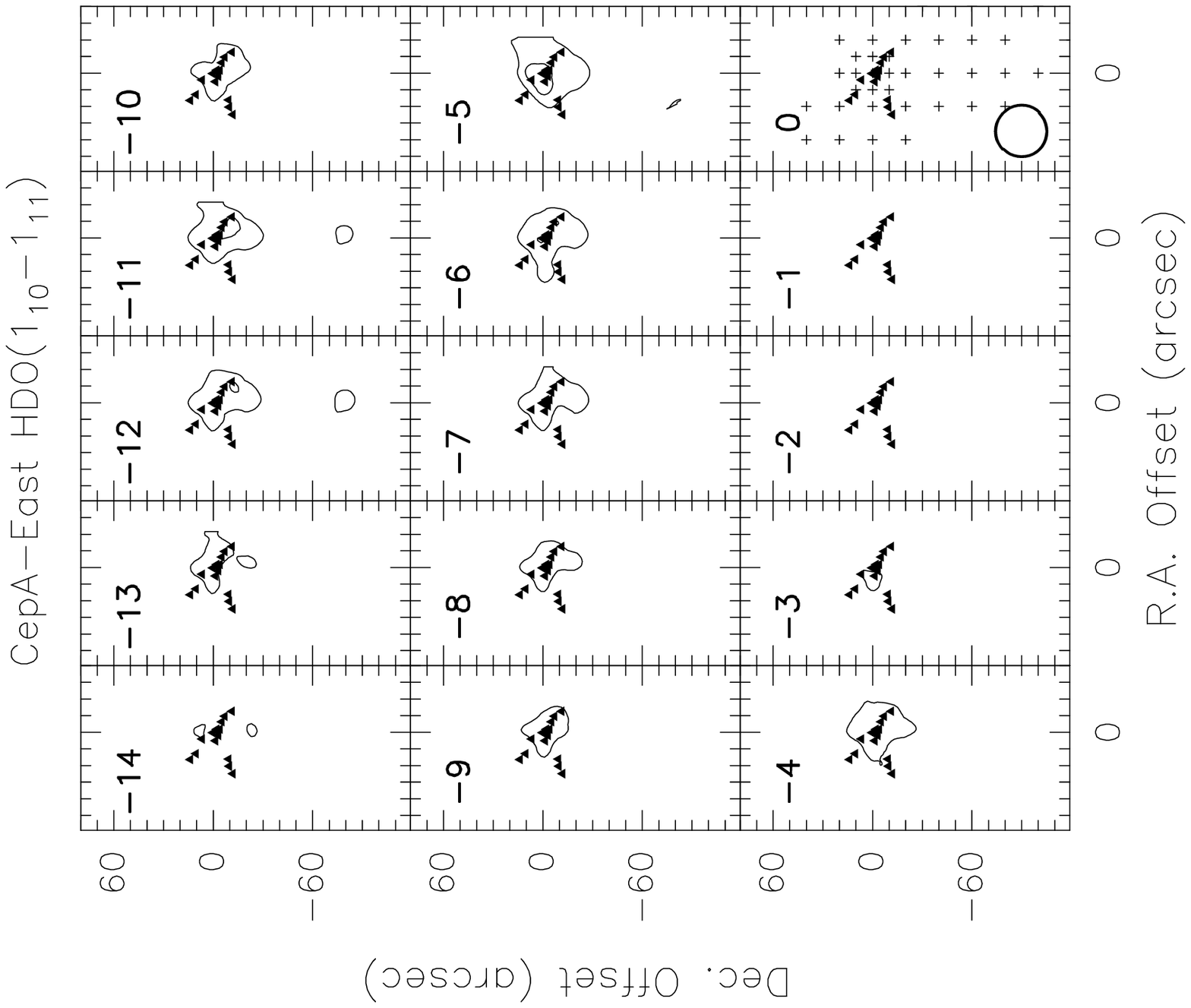,width=11cm,angle=-90}
\caption{Channel map of the
HDO $J_{\rm K_-K_+}$ = 1$_{\rm 10}$--1$_{\rm 11}$ emission towards
CepA-East. Each panel shows the emission integrated over a velocity interval
of 1 km s$^{-1}$ centred at the value given in the left corner.
Symbols are drawn as in Fig. 1.
The ambient velocity emission is --10.7 km s$^{-1}$ according
to CS $J$ = 5--4 and C$^{18}$O $J$ = 2--1 measurements (see text).
The contours range are 0.03 ($\sim$3$\sigma$) and 0.06
K km s$^{-1}$.}
\end{figure*}

\begin{figure*}
\psfig{file=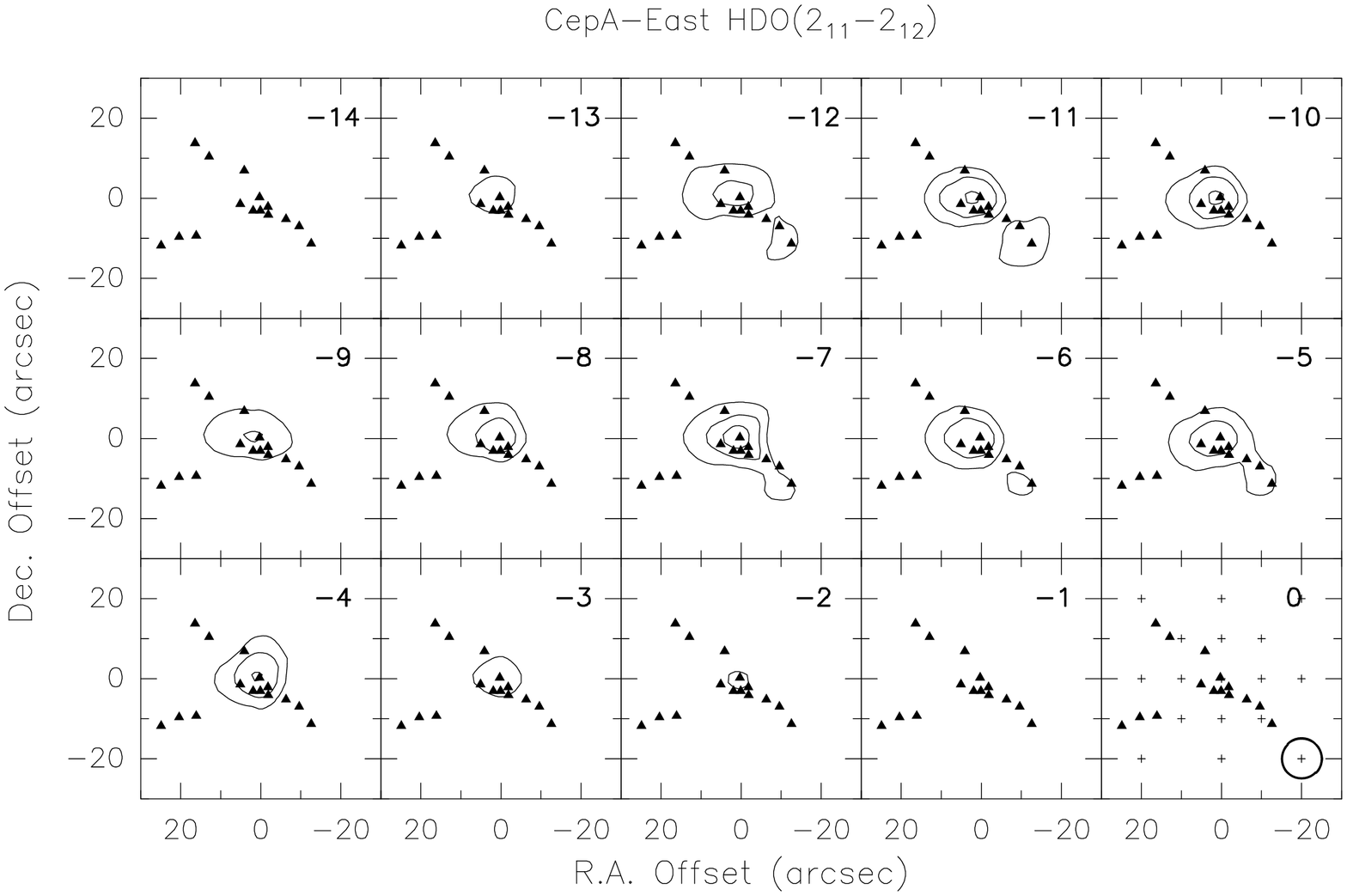,width=16cm,angle=-90}
\caption{Zoom of the channel map of the
HDO $J_{\rm K_-K_+}$ = 2$_{\rm 11}$--2$_{\rm 12}$ emission towards
CepA-East. Each panel shows the emission integrated over a velocity interval
of 1 km s$^{-1}$ centred at the value given in the left corner.
Symbols are drawn as in Fig. 1.
The ambient velocity emission is --10.7 km s$^{-1}$ according
to CS $J$ = 5--4 and C$^{18}$O $J$ = 2--1 measurements (see text).
The contours range are 0.12 ($\sim$3$\sigma$), 0.24, and 0.36
K km s$^{-1}$.}
\end{figure*}

Figure 5 shows the contour maps of the $J_{\rm K_-K_+}$ = 1$_{\rm 10}$--1$_{\rm 11}$, and
2$_{\rm 11}$--2$_{\rm 12}$ integrated HDO emission towards CepA-East. In particular, 
considering the CS results, the southern 
region has also been mapped to look for signposts of outflow motions.
The 1$_{\rm 10}$--1$_{\rm 11}$ emission clearly indicates (i) a structure elongated
along the SW direction with a peak near the centre of the map, 
and (ii) another elongated feature pointing
towards South (see Sect. 3.3). 
On the other hand, the 2$_{\rm 11}$--2$_{\rm 12}$ HDO map, obtained with higher 
resolution (10$\arcsec$), reveals a $\sim$15$\arcsec$ (0.05 pc) clump centered at the
HW2 coordinates, where the YSOs are located, and a second 
unresolved structure lying along the SW VLA jet.

In order to study the HDO kinematics, we present in Figures 6 and 7 the velocity 
channel maps of the 1$_{\rm 10}$--1$_{\rm 11}$ and 2$_{\rm 11}$--2$_{\rm 12}$ emissions,
respectively. From these maps we argue that
HDO in CepA-East traces the outflow activity as well as a small region
around the driving YSOs.
A quite complex scenario is therefore drawn:
(i) a central component, clearly shown by the 2$_{\rm 11}$--2$_{\rm 12}$ line, 
which is associated both with ambient velocity and 
higher redshifted (up to --2 km s$^{-1}$) velocities; (ii)
a component associated with the SW VLA chain and emitting in both HDO lines 
and (iii) a final component 
pointing towards South and detected through the 1$_{\rm 10}$--1$_{\rm 11}$ line.
These last two components emit at redshifted velocities (up to --4 km s$^{-1}$) and
at --12 and --11 km s$^{-1}$, i.e. at velocities close to the ambient one.
Note that the --12 and --11 km s$^{-1}$ panels of Fig. 6 seem to indicate 
also the possible occurrence of a HDO 1$_{\rm 10}$--1$_{\rm 11}$ clump
at (0$\arcsec$,--80$\arcsec$), along the southern direction, suggesting
a connection with the central feature pointing towards South. 
This southern emission has not been detected
with the 2$_{\rm 11}$--2$_{\rm 12}$ line and therefore for the sake of clarity in the 
corresponding channel maps in Fig. 7 only a zoom of a smaller region is shown.

\subsection[]{HDO and H$_2$S}

In order to further investigate the southern molecular outflow, the present
HDO maps can be compared with the H$_2$S ones 
reported by Codella et al. \shortcite{code03}. 
In particular, Fig. 8 shows the contour maps (thin line) 
of the H$_2$S(1$_{\rm 10}$-1$_{\rm 01}$) emission
integrated over the redshifted velocity interval typical
of the southern outflow: --7,--3 km s$^{-1}$. 
The HDO(1$_{\rm 10}$-1$_{\rm 01}$) map, integrated over the same 
velocity range, is drawn by using a thick contour.
From the H$_2$S contours it is possible to detect the SE outflow 
(ending at the position of the clump
A) and the southern outflow, which extends at the position of the clump C.
The HDO map is in agreement with the picture given by H$_2$S,   
both suggesting an outflow activity towards South.
This scenario could reflect different
chemical and/or physical conditions for the regions where
the different flows are moving. The southern outflow
seems chemically enriched with respect to the typical gas composition of
the dark clouds.
It shows a redshifted component (see Fig. 6), whereas the 
blueshifted counterpart could be tentatively singled out in the emission around $\sim$ --12 km s$^{-1}$,
i.e. at velocities close to the ambient one, but slightly blueshifted with respect to the --10.7
km s$^{-1}$ value. In particular, the tentative detection of HDO emission  
at (0$\arcsec$,--80$\arcsec$) shown in the channel maps in 
Fig. 6 is supported by the occurrence of a  
H$_2$S red- and blueshifted (see Fig. 4 of Codella et al. 2003) clump at the same position.

\begin{figure}
\psfig{file=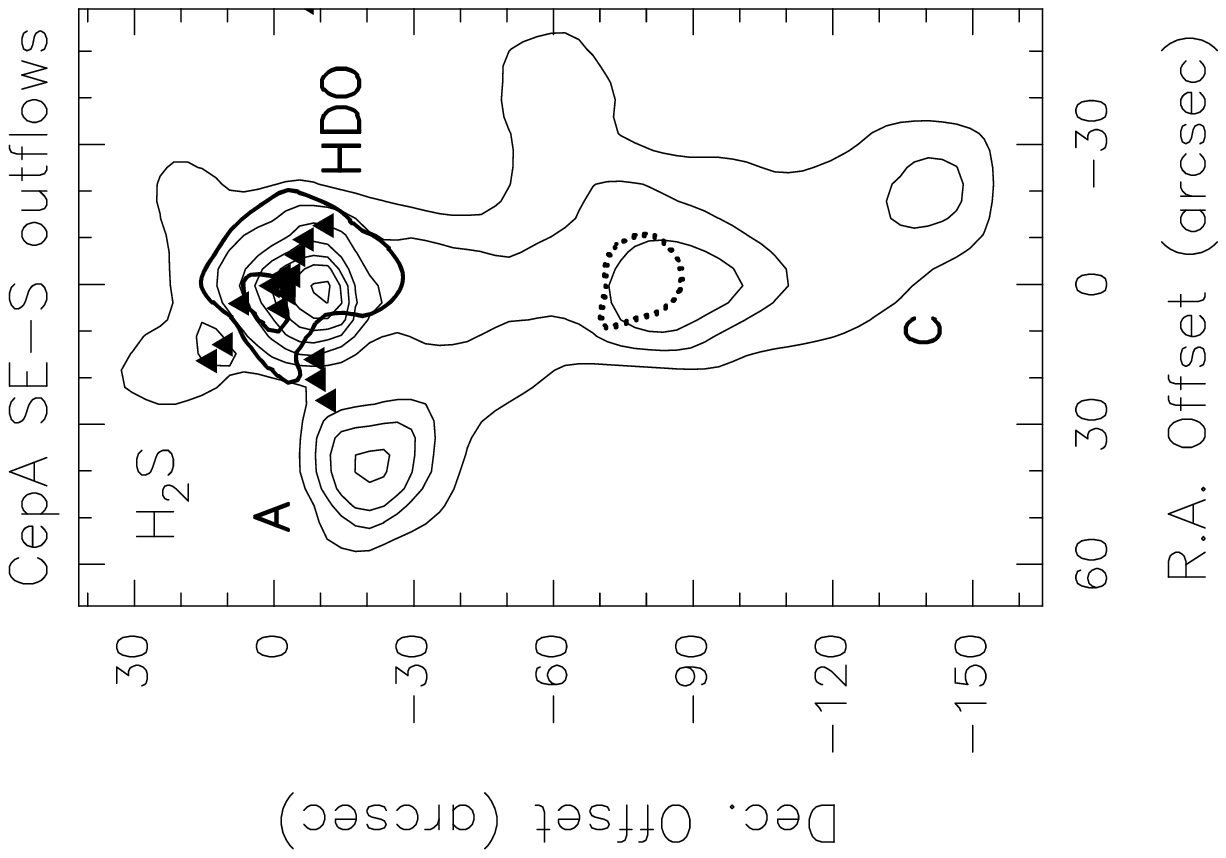,width=8cm,angle=-90}
\caption{Contour maps (thin line) of the H$_2$S(1$_{\rm 10}$-1$_{\rm 01}$) redshifted emission
(Codella et al. 2003) integrated
over the velocity interval --7,--3 km s$^{-1}$.
Two main flows are clearly drawn: one pointing to SE (and ending at the clump A position),
and another towards South (ending at clump C). Symbols are drawn as
in Fig. 1. The thick solid line shows
the contour maps of the HDO(1$_{\rm 10}$-1$_{\rm 01}$)
emission integrated over the same range of velocity.
The thick dotted line shows the distribution of the
HDO(1$_{\rm 10}$-1$_{\rm 01}$) emission integrated over the
interval --13,--11 km s$^{-1}$ and tentatively points out a
blueshifted counterpart (see the channel maps of Fig. 6).}
\end{figure}

\begin{figure}
\psfig{file=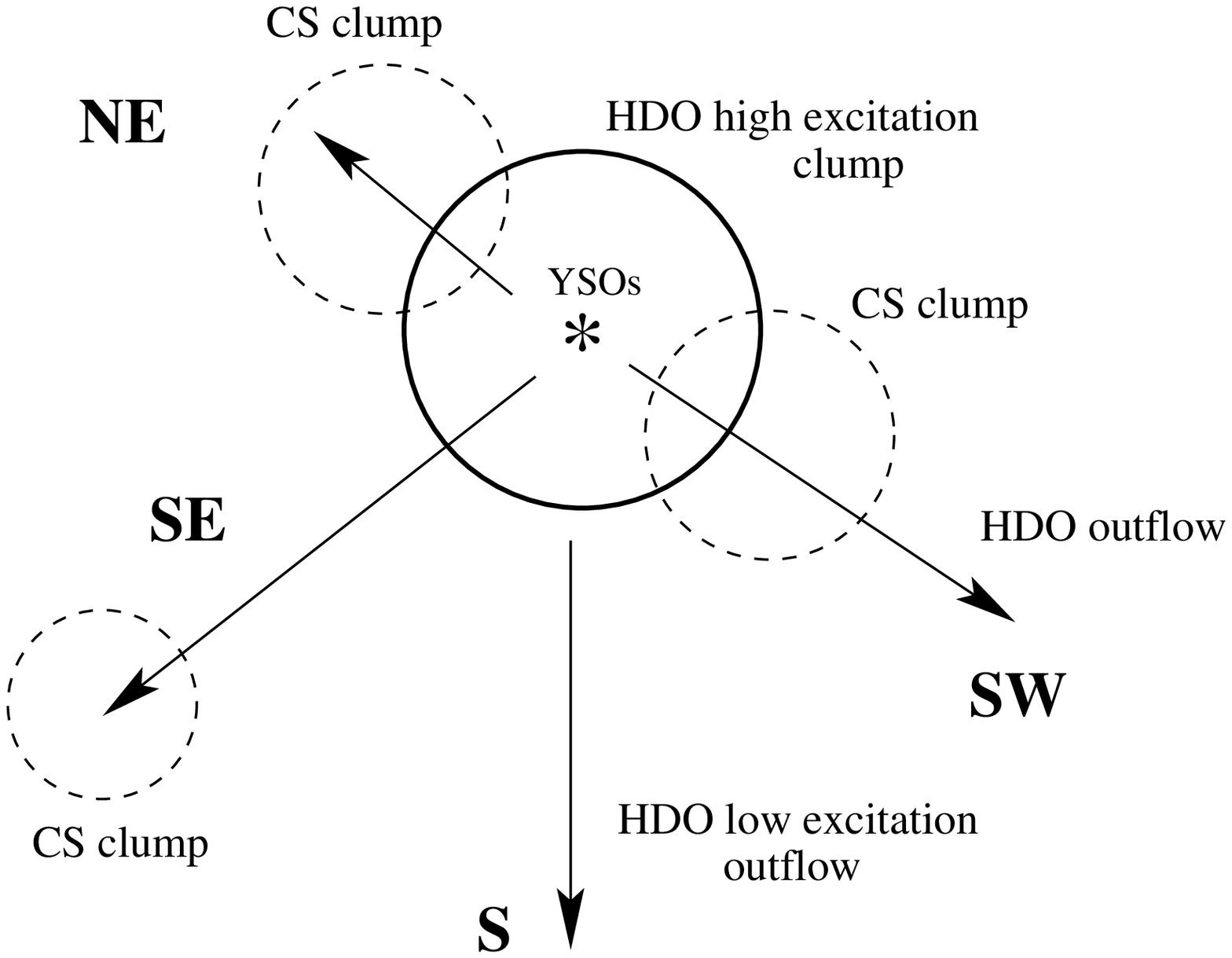,width=8.5cm,angle=-90}
\caption{A schematic picture (not to scale) of the directions of the multiple outflows
driven by the CepA-East YSOs as traced by HDO and CS emission (see text). Following the
channel maps four main flows are identified: three (SW, NE, and SE) associated with the
VLA shocked chains and with CS clumps, and one (S) traced by the HDO(1$_{\rm 10}$-1$_{\rm 01}$)
emission. The higher excitation HDO(2$_{\rm 11}$--2$_{\rm 12}$) line well defines
the SW outflow and with the central region close to the YSOs coordinates (marked by an asterisk).}
\end{figure}

In Fig. 9 we summarise all the information given by the CS and HDO channel maps,
drawing a schematic picture with the main clumps and outflow directions.
The high excitation HDO(2$_{\rm 11}$--2$_{\rm 12}$)
line is tracing the region close to the YSOs, marked in Fig. 9 by an asterisk
at (0$\arcsec$,0$\arcsec$) offset representing the HW2 coordinates.
Four main flows are identified: three (SW, NE, and SE) associated with the VLA sources
and probably accelerating high density material traced by CS.
The SW outflow is traced also by HDO emission.
In addition HDO reveals a fourth outflow pointing
towards South and not traced by the VLA sources nor by the CS structures.
Finally, we note that the highest HDO excitation conditions
occur near the YSOs, where the beam deconvolved line intensity ratio
$R_{\rm HDO}$ $\simeq$ $T_{\rm mb}$(2$_{\rm 11}$--2$_{\rm 12}$)/$T_{\rm mb}$(1$_{\rm 10}$-1$_{\rm 01}$)
is about 2. On the other hand, the southern flow suggests lower excitation conditions,
with $R_{\rm HDO}$ $\simeq$ 0.6 at the (0$\arcsec$,--20$\arcsec$) offset.

\section[]{A spectral survey towards CepA-East}

\begin{figure}
\psfig{file=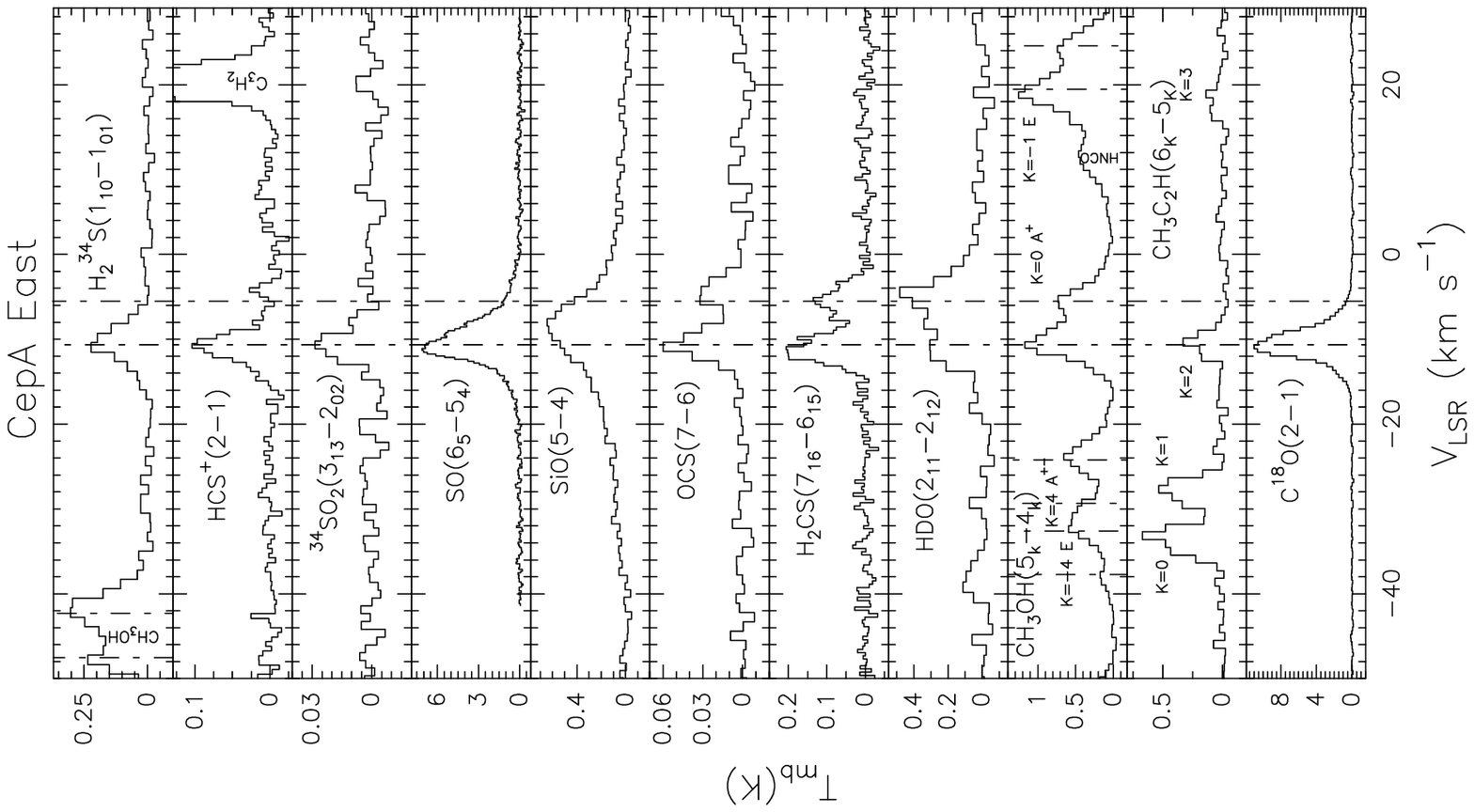,width=8cm,angle=-90}
\caption{
Molecular line profiles observed towards CepA East: species and transitions are reported.
The dashed lines stand for the ambient LSR velocity (--10.65 km s$^{-1}$), according to the
C$^{18}$O measurement, and for the component at --5.51 km s$^{-1}$ well
outlined e.g. in the H$_2$CS profile.
In case of the CH$_3$OH(5$_{\rm K}$--4$_{\rm K}$) spectrum, four hyperfine lines at different
excitations are present: $K$=0 A$^+$ ($E_{\rm u}$=35 K), $K$=--1 E ($E_{\rm u}$=39 K),
$K$=--4 E ($E_{\rm u}$=122 K), and $K$=4 A$^{+-}$ ($E_{\rm u}$=115 K).
The CH$_3$C$_2$H(6$_{\rm K}$--5$_{\rm K}$) pattern shows the $K$=0,1,2,3 lines.
The small vertical labels CH$_3$OH
($E_{\rm u}$=126 K), C$_3$H$_2$, and
HNCO refers to three serendipity detections (see Table 1).
}
\end{figure}

\begin{table*}
\caption[] {Summary of the properties of the profiles observed towards CepA-East}
\begin{tabular}{cc}
\hline
\multicolumn{2}{c}{Molecular tracers in CepA-East} \\
\hline
CH$_3$OH, HCS$^+$, HDO, H$_2$CS, OCS \hspace{1.05cm} $\Longrightarrow$ & Ambient, Outflow, and Peak @ --5.5 km s$^{-1}$
\\
CO, CS, H$_2$S, SiO, SO, SO$_2$ \hspace{1.98cm} $\Longrightarrow$ & Ambient and Outflow \\
C$^{13}$CH, C$_3$H$_2$, CH$_2$CO, CH$_3$C$_2$H, HC$^{18}$O$^+$ $\Longrightarrow$ & Only Ambient \\
\hline
\end{tabular}
\end{table*}

Figure 10 presents the most representative examples of line profiles observed
towards CepA-East.
Different molecules exhibit different spectral behaviours and they can be grouped in three
classes, summarised in Table 3: (i) ambient species 
(C$^{13}$CH, C$_3$H$_2$, CH$_2$CO, CH$_3$C$_2$H, HC$^{18}$O$^+$), not detected at high velocities
and with relatively narrow (3-4 km s$^{-1}$) lines
at velocities close to the ambient one (hereafter called ambient peak), 
(ii) outflow tracers  
(CO, CS, H$_2$S, SiO, SO, SO$_2$), which show extended wings 
(e.g. $v$--$v_{\rm LSR}$ up to $\sim$ 30 km s$^{-1}$ for CO and 10 km s$^{-1}$ for SO) 
and span the whole range of observed velocities, and (iii) species 
(OCS, H$_2$CS, HDO, and CH$_3$OH) which are associated with wings and, in addition, 
show a redshifted secondary peak at
--5.5 km s$^{-1}$ (hereafter called outflow peak), well separated (by about 5 km s$^{-1}$) from the
ambient velocity. 
Unfortunately, the present data do not allow us to clarify the spatial 
distribution of the spectral outflow peak, i.e. whether it is
tracing a small clump or it is related to a more extended
structure. However, the comparison between HDO and H$_2$S emissions
shown in Fig. 9 clearly suggests that the spectral outflow peak is tracing an elongated 
structure flowing towards South. Only accurate CH$_3$OH, H$_2$CS, and OCS maps
can confirm this conclusion.

Figure 11 displays
the CH$_3$OH(5$_{\rm K}$--4$_{\rm K}$) and (3$_{\rm K}$--2$_{\rm K}$) spectra in the
upper and lower panel, respectively: the dashed lines
indicate the predicted positions of the hyperfine component at different excitations. 
Although different lines are blended, from these
spectacular emissions it is possible to see that almost all the components, including
those at very high excitations with $E_{\rm u}$ $\simeq$ 120 K, 
have been detected and show the redshifted peak. 
It is worth noting that other very high excitation CH$_3$OH lines 
(up to $E_{\rm u}$ = 390 K) have been serendipitously detected, as reported in Table 1.
Finally, Fig. 12 shows the striking differences observed on the
profiles of several transitions of HDO (upper panels) and H$_2$CS (lower panels).
From these spectra, it is clear that 
the two HDO peaks are associated with roughly similar excitation conditions, whereas 
for H$_2$CS, the --5.5 km s$^{-1}$ peak increases its intensity with 
respect to the --10.7 km s$^{-1}$ peak with excitation indicating different physical conditions.
For instance, the ratio between the mean brightness temperatures of the outflow and ambient
peaks is $\sim$ 0.6 for the $J_{\rm K_-K_+}$ = 7$_{16}$--6$_{15}$ transition at 60 K,
while it is $\le$ 0.1 for the $J_{\rm K_-K_+}$ = 3$_{13}$--2$_{12}$ line at 23 K.

\begin{figure}
\psfig{file=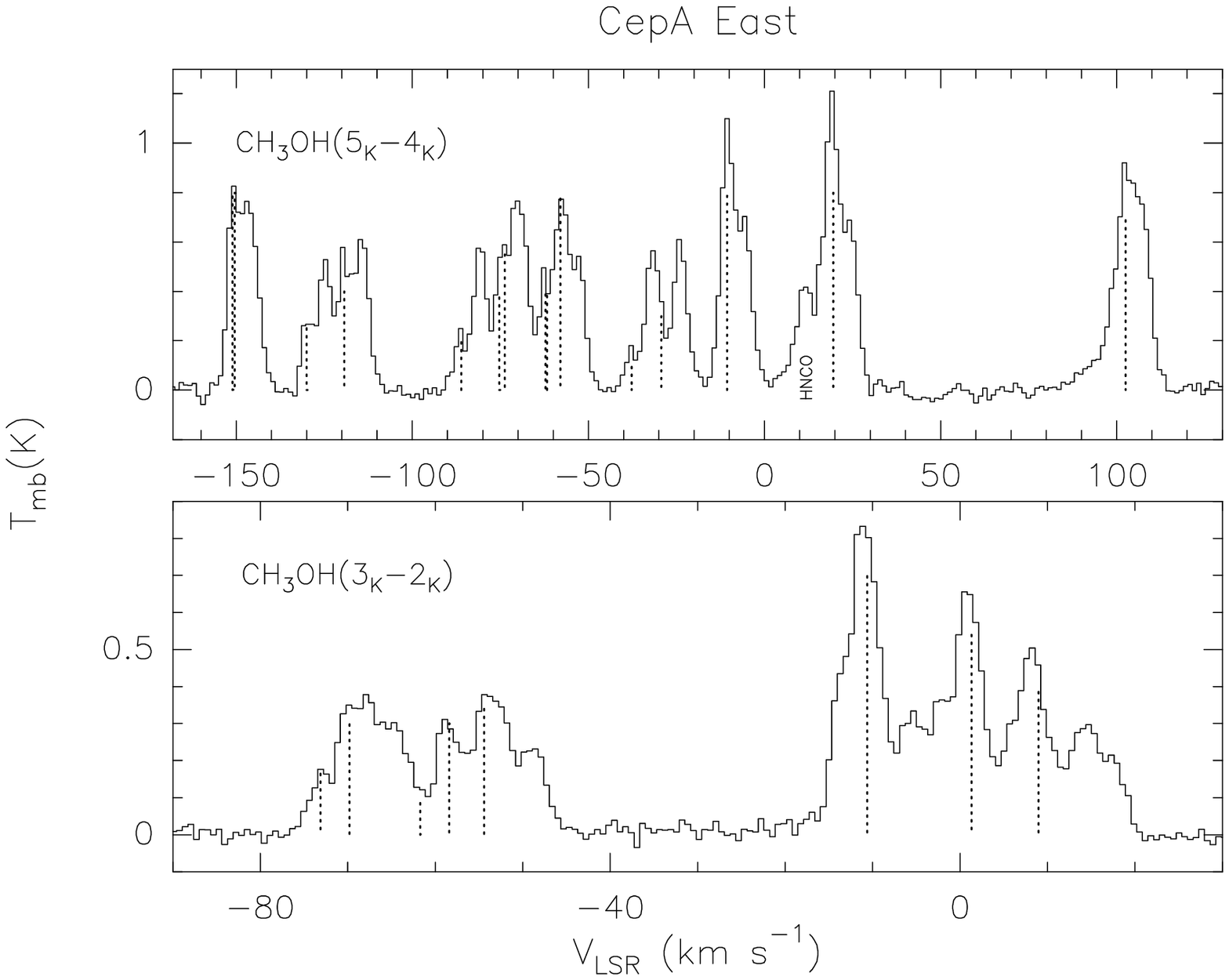,width=8.3cm,angle=-90}
\caption{
CH$_3$OH(5$_{\rm K}$--4$_{\rm K}$) and CH$_3$OH(3$_{\rm K}$--2$_{\rm K}$) line profiles observed
towards CepA East.
The dotted lines stand for the predicted positions of the hyperfine components at different
excitations.
The HNCO(11$_{\rm 011}$--10$_{\rm 010}$) line is also marked in the methanol (5$_{\rm K}$--4$_{\rm
K}$)
pattern. The velocity scale of the spectra is calculated with respect to the frequency of the
5$_{\rm 0}$--4$_{\rm 0}$A$^+$ (241791.43 MHz)  and 3$_{\rm 0}$--2$_{\rm 0}$A$^+$ (145103.23 MHz)
lines.}
\end{figure}

\begin{figure}
\psfig{file=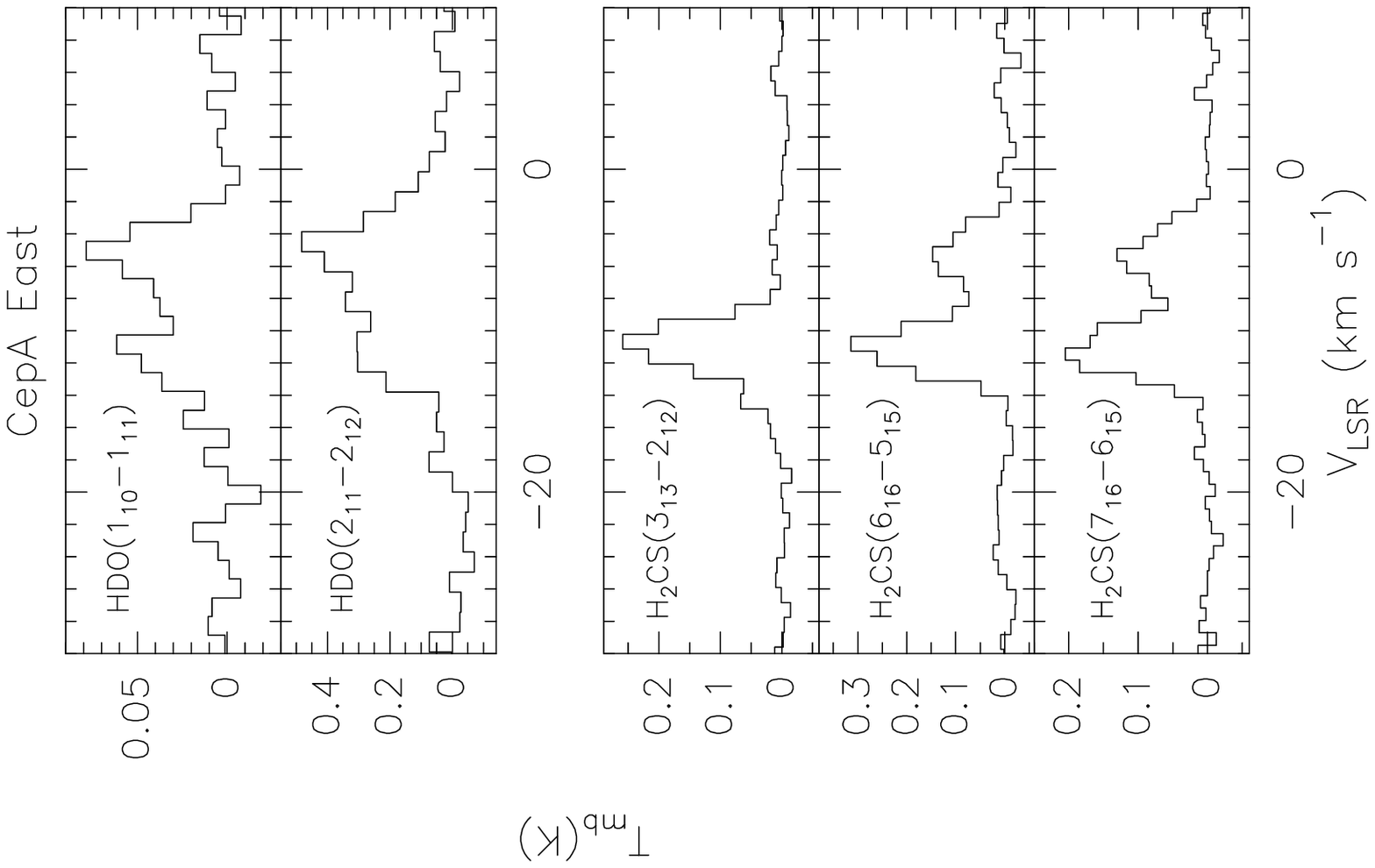,width=8cm,angle=-90}
\caption{
HDO (upper panels) and H$_2$CS (lower panels) 
line profiles observed towards CepA East: transitions are reported (see Table 1).
Note how for H$_2$CS the --5.5 km s$^{-1}$ component increases its intensity
with respect the --10.7 km s$^{-1}$ peak with excitation.} 
\end{figure}

In conclusion, the obtained line profiles indicate that at the CepA-East positions the emissions
due to the clump hosting the YSOs and to the associated molecular outflows coexist and produce
two distinct line peaks which suggest different excitation conditions. 
The results suggest also that different molecules can trace different excitation
conditions at the same observed velocity.
This gives us a precious opportunity to perform a multiline analysis in order to clarify 
the physical conditions associated with the ambient emission and those associated
with the southern chemically rich molecular outflow.

\section[]{Derived gas parameters}

\subsection[]{Analysis procedures}

By means of statistical-equilibrium calculations and using the four observed SO lines,
it is possible to estimate the total column density ($N_{\rm tot}$) as well as the kinetic
temperature ($T_{\rm k}$) and the hydrogen density ($n_{\rm H_2}$). A Large Velocity Gradient (LVG)
model and the collisional rates from Green \shortcite{green} have been used.
Due to the lack of SO maps, these calculations have been performed using the mean-beam
brightness temperatures relative to the intensities of the four spectra, without beam filling
factor correction.
If the source was definitely smaller than the four beamwidths, the LVG calculations
would lead to an overestimate of the column densities (up to a factor 9) and of excitation conditions
(up to factors 1.2 and 2 for $T_{\rm k}$ and $n_{\rm H_2}$, respectively).

\begin{figure}
\psfig{file=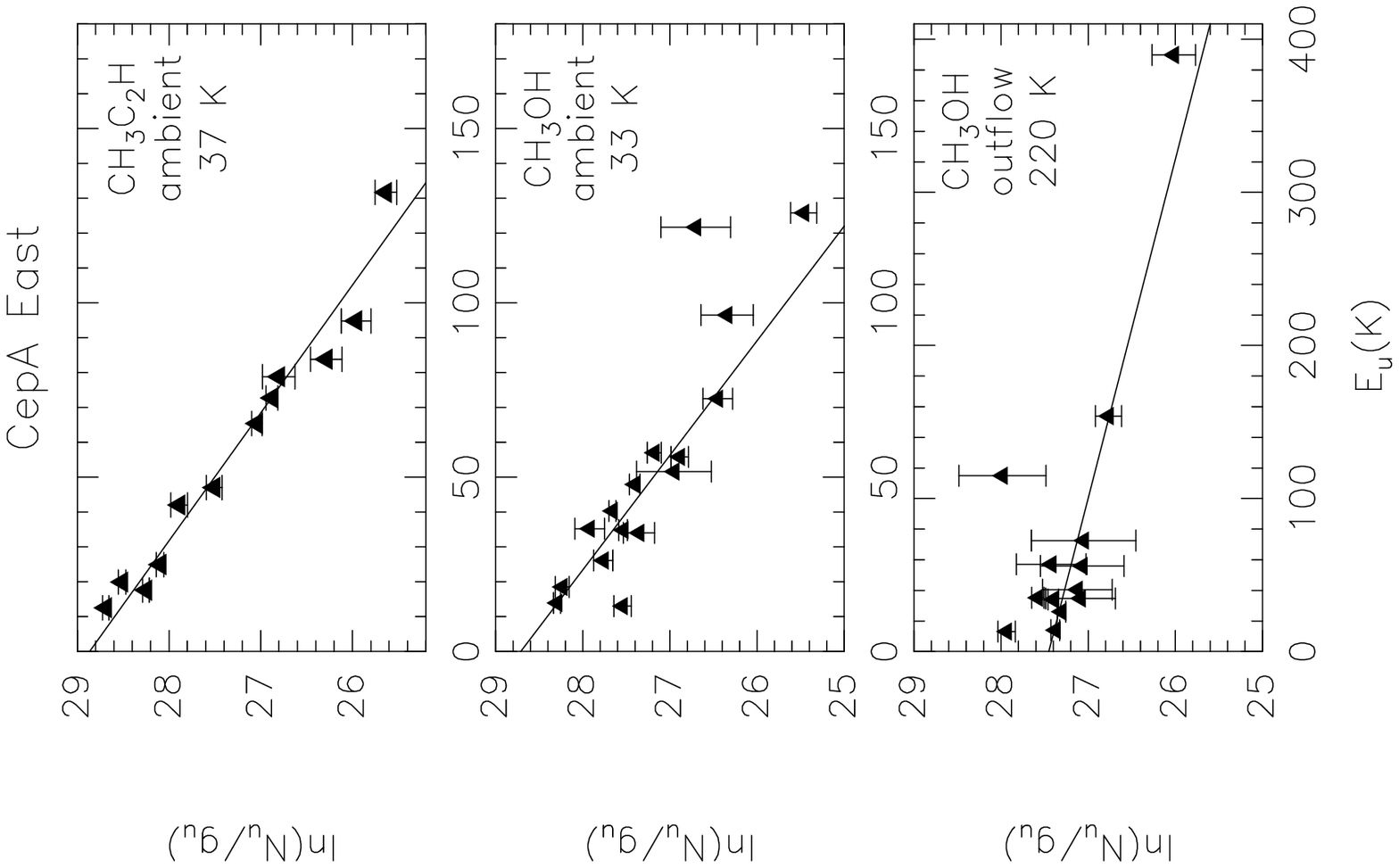,width=8cm,angle=-90}
\caption{Rotation diagrams for the CH$_3$C$_2$H (upper panel) and CH$_3$OH
(middle and lower panels) transitions measured towards CepA-East. For the methanol emission
two spectral regimes have been considered: ambient and outflow (see text).
The parameters $N_{\rm u}$, $g_{\rm u}$, and $E_{\rm u}$ are, respectively, the column density,
the degeneracy and the energy for the upper levels of the transitions reported.
The derived values of the rotational temperature are reported.}
\end{figure}

\begin{figure}
\psfig{file=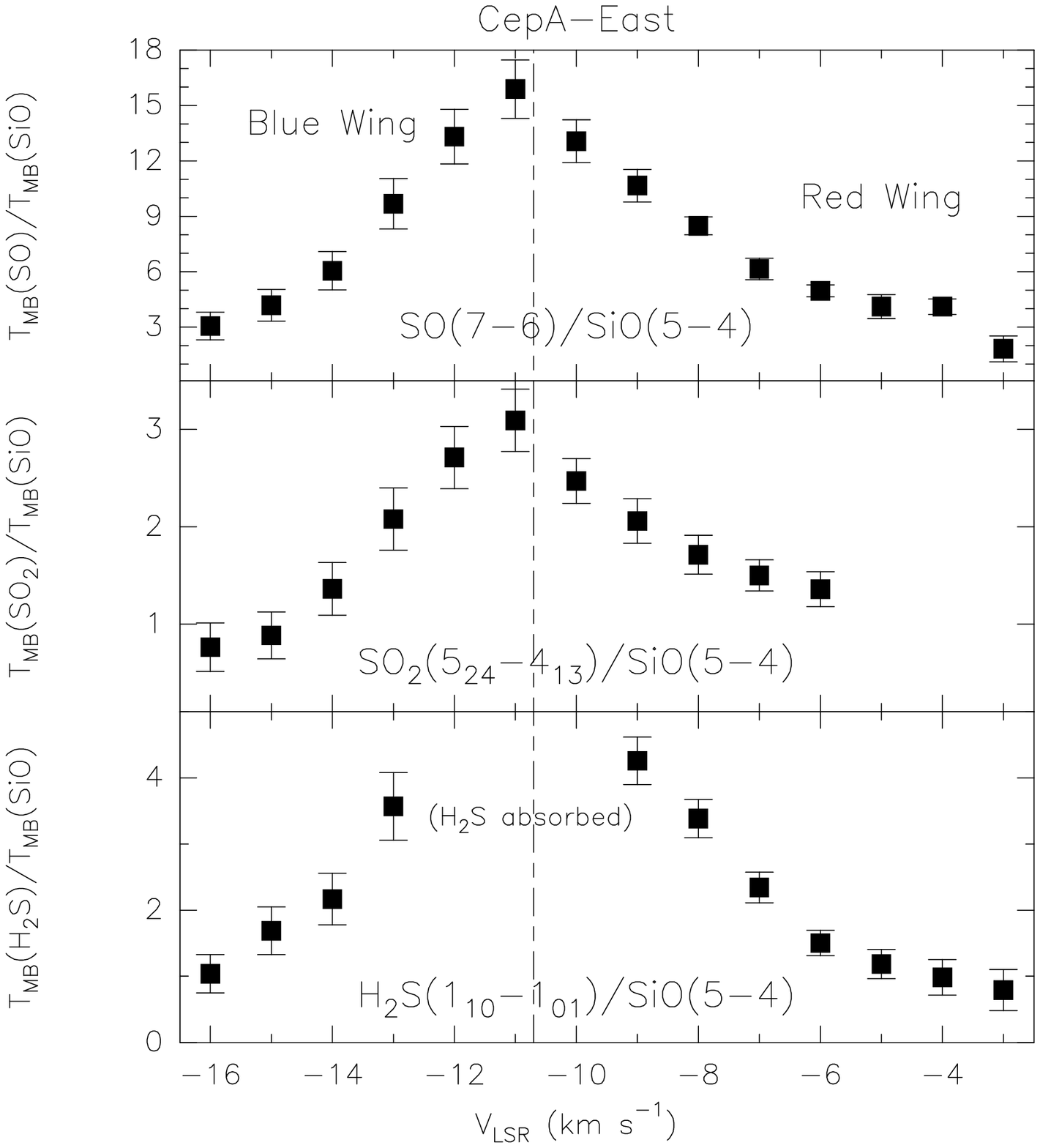,width=8cm,angle=-90}
\caption{Distribution with velocity of the ratio between the brightness temperatures of the
SO(7--6) ($E_{\rm u}$ = 35 K; upper panel), H$_2$S(1$_{10}$--1$_{01}$) ($E_{\rm u}$ = 28 K; middle
panel),
SO$_2$(5$_{24}$--4$_{13}$) ($E_{\rm u}$ = 24 K; lower panel) lines and of the SiO(5--4) ($E_{\rm u}$
=
21 K)
profile as detected towards CepA-East. The H$_2$S and SO$_2$ data are taken from
Codella et al. (2003). The HPBWs of the observations used here are: 9$\arcsec$ (SO),
10$\arcsec$ (SO$_2$), and 11$\arcsec$ (H$_2$S, SiO).
The dot-dashed line underlines
the ambient LSR velocity according to the CS $J$ = 5--4 and C$^{18}$O
$J$ = 2--1 measurement.}
\end{figure}

Following Cesaroni et al. \shortcite{cesa91}, statistical equilibrium computations in the 
LVG approximation have been used to analyse the CS and C$^{34}$S spectra.
In particular, for optically thin conditions, the line brightness temperatures depend only
on density and, to a minor degree, on temperature. In this case, we excluded CS(2-1) which
clearly shows self-absorption effects and we used CS(5-4) as well as the C$^{34}$S lines
to derive $n_{\rm H_2}$ estimates.

For the CH$_3$OH, CH$_3$C$_2$H, H$_2$CS, and $^{34}$SO$_2$ molecules,  
observed through at least three lines, in order to estimate the rotational 
temperature and column densities, the standard rotation diagram method,
assuming LTE and optically thin conditions, has been used.
For the other molecular species, the total column densities have been calculated using
the constants given in the databases for molecular spectroscopy
and considering the temperature
estimates derived from the LVG and rotation diagram results. 
For comparison, also the values calculated by using the H$_2$S and SO$_2$ emissions
(Codella et al. 2003) are reported.

When possible, different column densities have been derived for
the different lines at --10.7 and --5.5 km s$^{-1}$.
In particular, the methanol patterns clearly allow us to derive
different excitation conditions, while H$_2$CS
leads to an estimate of the temperature for the --10.7  km s$^{-1}$ component.
Finally, following the SO$_2$ analysis of Codella et al. \shortcite{code03}, we have
assumed two components also for CS and SO which show strong redshifted wing emission with
an intense emission around --5 km s$^{-1}$.

\subsection[]{Results}

The derived parameters have been summarised in Table 4.
The LVG results based on the SO spectra lead to kinetic temperatures of 60-100 K and
$n_{\rm H_2}$ $\sim$ 5 10$^6$ cm$^{-3}$ for the --10.7 km s$^{-1}$ line at
ambient velocity, while slightly higher temperatures (70-180 K) and definitely
higher densities (2 10$^6$-6 10$^7$ cm$^{-3}$) have been derived for the
--5.5 km s$^{-1}$ component. The total SO column densities are around 
4 10$^{14}$ and 5 10$^{13}$ cm$^{-2}$ for the ambient and outflow components,
respectively.

The CS LVG results indicate 
hydrogen number densities larger than 6 10$^4$ cm$^{-3}$
for the ambient component. For
the outflow component, the CS analysis leads to densities in the range
between 4 10$^3$ and 5 10$^4$ cm$^{-3}$.
On the other hand, the C$^{34}$S data for the ambient component
suggest lower densities with values
around 2 10$^4$ cm$^{-3}$. The C$^{34}$S emission is optically thinner than the CS one, and
in fact CS shows self-absorption whereas C$^{34}$S shows Gaussian
profiles; thus, it is reasonable to expect that C$^{34}$S is tracing inner regions
with respect to CS. In this case, the LVG results seem to give puzzling results 
for CepA-East with a high density envelope and a lower density central region.
One possible solution comes from the recent results reported by Bottinelli \& Williams
\shortcite{botti}, who analyse the large scale dynamics of CepA-East by using
a density profile measured from a 850 $\mu$m map of the region. 
The model proposed by these authors has a high density center with depleted CS 
and an outer envelope with a low $n_{\rm H_2}$ and normal CS abundance.
Therefore, our hydrogen density estimates could be biased by the fact that
the CS abundance is varying along the central region.

\begin{table*}
\caption[] {Column densities, temperatures, and hydrogen densities for the ambient and outflow
gas components}
\begin{tabular}{l|cccc|lccc|c}
\hline
\multicolumn{1}{c|}{ } &
\multicolumn{4}{c|}{AMBIENT} &
\multicolumn{4}{c|}{OUTFLOW PEAK} &
\multicolumn{1}{c}{ } \\
\multicolumn{1}{c|}{Molecules} &
\multicolumn{1}{c}{$N_{\rm tot}$$^a$} &
\multicolumn{1}{c}{$T_{\rm kin}$} &
\multicolumn{1}{c}{$n_{\rm H_2}$} &
\multicolumn{1}{c|}{$X[\hspace{0.2cm}]/X[H_2]$$^a$} &
\multicolumn{1}{c}{$N_{\rm tot}$$^a$} &
\multicolumn{1}{c}{$T_{\rm kin}$} &
\multicolumn{1}{c}{$n_{\rm H_2}$} &
\multicolumn{1}{c|}{$X[\hspace{0.2cm}]/X[H_2]$$^a$} & 
\multicolumn{1}{c}{$f$$^b$} \\
\multicolumn{1}{c|}{} &
\multicolumn{1}{c}{(cm$^{-2}$)} &
\multicolumn{1}{c}{(K)} &
\multicolumn{1}{c}{(cm$^{-3}$)} &
\multicolumn{1}{c|}{} &
\multicolumn{1}{c}{(cm$^{-2}$)} &
\multicolumn{1}{c}{(K)} &
\multicolumn{1}{c}{(cm$^{-3}$)} &
\multicolumn{1}{c|}{ } & 
\multicolumn{1}{c}{ } \\
\hline
C$^{13}$CH   & 0.6-3 10$^{13}$ & -- & -- & 0.2-3 10$^{-8}$$^c$ &  -- & -- & -- & -- & -- \\
C$_3$H$_2$   & 0.6-5 10$^{13}$ & -- & -- & 0.2-5 10$^{-10}$ &  -- & -- & -- & -- & -- \\
CH$_2$CO     & 0.3-1 10$^{15}$ & -- & -- & 8 10$^{-10}$-1 10$^{-8}$ &  -- & -- & -- & -- & -- \\
CH$_3$C$_2$H & 1 10$^{15}$ & 37 & -- & 0.3-1 10$^{-8}$ & -- & -- & -- & -- & -- \\
CH$_3$OH     & 7 10$^{14}$ & 33 & -- & 2-7 10$^{-9}$ & 3 10$^{15}$ & 220 & -- & 6 10$^{-8}$ & 20  \\
CO, C$^{18}$O & 1-4 10$^{19}$ & -- & -- & -- & 5 10$^{18}$ & -- & -- & -- & -- \\
CS, C$^{34}$S & 1-5 10$^{14}$ & -- & $>$ 6 10$^4$ & 0.3-5 10$^{-9}$ & 3 10$^{13}$ & -- & 4 10$^3$-5
10$^4$ & 6 10$^{-10}$ & 1 \\
HC$^{18}$O$^+$ & 1-6 10$^{12}$ & -- & -- & 3 10$^{-8}$-1 10$^{-10}$$^c$ & -- & -- & -- & -- & -- \\
HCS$^+$       & 0.3-1 10$^{13}$ & -- & -- & 8 10$^{-12}$-1 10$^{-10}$ & 3 10$^{12}$ & -- & -- & 6
10$^{-11}$ & 4 \\
HDO           & 2-5 10$^{14}$ & -- & -- & 0.5-5 10$^{-9}$ & 6 10$^{14}$ & -- & -- & 1 10$^{-8}$ & 11 \\
H$_2$CS       & 3 10$^{13}$ & 20 & -- & 0.8-3 10$^{-10}$ & 1 10$^{14}$ & -- & -- & 2 10$^{-9}$ & 16 \\
H$_2$S, H$_2$$^{34}$S$^{d,e}$ & 5 10$^{14}$ & 27 & -- & 1-5 10$^{-9}$ & --$^d$ & --$^d$ & --$^d$ &
--$^d$ &   --$^d$ \\
HNCO                & 2-3 10$^{13}$ & -- & -- & 0.5-3 10$^{-10}$ & -- & -- & -- & -- & -- \\
OCS                   & 2-5 10$^{13}$ & -- & -- & 0.5-5 10$^{-10}$ & 5 10$^{13}$ & -- & -- & 1
10$^{-9}$ & 11  \\
SiO$^d$  & 1 10$^{13}$ & 150    &  2 10$^6$ & 0.3-1 10$^{-10}$ &  --$^d$ & --$^d$ & --$^d$ & --$^d$ & --$^d$ \\
SO     & 4 10$^{14}$ & 60-100 & 4-6 10$^6$ & 1-4 10$^{-9}$ & 5 10$^{13}$ & 70-180 & 2 10$^6$-6 10$^7$
& 1 10$^{-9}$ & 0.7 \\
SO$_2$$^e$ & 3 10$^{15}$ & 75 & -- & 0.9-3 10$^{-8}$ & 1 10$^{15}$ & 130 & -- & 2
10$^{-8}$ & 1.5 \\
$^{34}$SO$_2$ & 1 10$^{14}$ & 200 & -- & 0.3-1 10$^{-9}$ & -- & -- & -- & -- & -- \\
\hline
\end{tabular}
\begin{center}
$^a$ The column densities and
the abundances regarding the species detected through only one line have
been derived assuming $X$[CO]/$X$[H$_2$] = 10$^{-4}$,
and $T_{\rm kin}$ = 20-100 K (ambient peak) and $T_{\rm kin}$ = 200 K
(outflow peak), following the temperature measurements based on the CH$_3$C$_2$H,
CH$_3$OH, SO, and SO$_2$ emissions.
$^b$ Average ratio between the outflow and the ambient abundances. 
$^c$ For C$^{13}$CH and
HC$^{18}$O$^+$, the abundances refer to the most abundant isotopomers (CCH and HCO$^+$) calculated by
assuming $^{12}$C/$^{13}$C = 89 and $^{16}$O/$^{18}$O = 490 (Wilson \& Rood 1994).
$^d$ For H$_2$S and SiO the ambient and outflow components cannot be disentangled; 
thus the values refer to the whole emission and the abundances are not reported.
$^e$ The H$_2$S and SO$_2$ results are taken from the Codella et al. (2003) paper: the H$_2$$^{34}$S
emission has been used to derive the optical depth and consequently refine the H$_2$S column density.
\end{center}
\end{table*}

The least-square fits to all detected lines in the CH$_3$OH rotation diagram,
shown in Fig. 13, gives, 
for the ambient component (middle panel), rotational temperatures of about 30 K, while the
total methanol column density is about 7 10$^{14}$ cm$^{-2}$.
The CH$_3$OH temperature is confirmed by the CH$_3$C$_2$H rotation diagram which leads to 37 K.
This suggests that at the ambient velocities
both molecules trace similar environments.
On the other hand, for the outflow component (lower panel of Fig. 13) the
derived temperature is 220 K and the CH$_3$OH column density is
3 10$^{15}$ cm$^{-2}$. However, due to the
blending between ambient and outflow components for the methanol spectral patterns,
the plot for the outflow component is less clear
and the fit is definitely more doubtful. In any case, the comparison between
the plots of the ambient component and the outflow one, which shows high excitation  
($E_{\rm u}$ = 390 K) CH$_3$OH emission, clearly indicates a definitely higher temperature 
for the outflow component.

Using the H$_2$CS and $^{34}$SO$_2$ emissions, 
an estimate of the rotational temperature of
the ambient emission, observed in three lines, has also been obtained. 
The results lead to low temperatures (20 K)
for H$_2$CS and definitely higher values (200 K) for $^{34}$SO$_2$,
suggesting that the H$_2$CS spectral peak at rest velocity mainly traces
relatively cool ambient material. The
column densities are $N_{\rm H_2CS}$ $\simeq$ 3 10$^{13}$  and
$N_{\rm ^{34}SO_2}$ $\simeq$ 10$^{14}$ cm$^{-2}$.

Taking into account the temperature estimates obtained from the SO and CH$_3$OH data,
the total column densities for the molecular species observed in one or two
lines have been calculated using the spectroscopic constants
given in the literature and summarised in the JPL (Pickett et al. 1998), 
NIST (Lovas 2004), Cologne (M\"uller et al. 2001), and Leiden (Sch\"oier et al. 2005) databases and 
considering a kinetic temperature in the range 20-100 K 
for the ambient component and 200 K for the outflow one.
Table 4 reports the values of the total column densities for
the ambient emission: $N_{\rm C^{13}CH}$ and
$N_{\rm C_3H_2}$ are about 1 10$^{14}$ cm$^{-2}$, $N_{\rm HNCO}$ and $N_{\rm OCS}$
$\simeq$ 3 10$^{13}$ cm$^{-2}$, $N_{\rm CH_2CO}$ $\sim$ 5 10$^{15}$ cm$^{-2}$,
$N_{\rm HC^{18}O^+}$ and $N_{\rm HCS^+}$ $\sim$ 3--5 10$^{12}$ cm$^{-2}$, 
while $N_{\rm HDO}$ is around 3 10$^{14}$ cm$^{-2}$.
On the other hand, for the outflow component we have column densities around
3 10$^{12}$, 6 10$^{14}$, and 5 10$^{13}$ cm$^{-2}$ for HCS$^+$, HDO, and
OCS, respectively.
We note that the HCS$^+$/CS abundance ratio is quite large, $\sim$ 0.1, in the outflow peak, 
close to that measured in dark clouds (Ohishi et al. 1992),
suggesting that the HCS$^+$ emission has been overestimated, due the low S/N ratio
of the HCS$^+$ outflow peak, and/or that the derived CS abundance is 
underestimated, probably due to the
blending between ambient and outflow spectral components (Sect. 5.1). 

The SiO results shown in Table 4 refer to the multiline SiO survey given by Codella
et al. \shortcite{codesio}, whereas the H$_2$S and SO$_2$ results are 
taken from the Codella et al. \shortcite{code03} paper: the H$_2$$^{34}$S
emission have been used here to derive the optical depth, 
$\tau$$_{(1_{\rm 10}-1_{\rm 01})}$ $\simeq$ 1--2,
and consequently refine the H$_2$S column density.
Finally, it is worth noting that 
for H$_2$S and SiO the ambient and outflow components cannot be disentangled
both showing continuous extended wings, and
thus the values in Table 4 refer to the whole emission.

In conclusion, we find that in CepA-East
at ambient velocities (i) the gas is associated with high densities, $>$ 10$^5$ cm$^{-3}$,
and (ii) different components at different temperatures coexist, ranging from the
relatively low kinetic temperatures, less than 50 K, 
measured with H$_2$S, CH$_3$OH, H$_2$CS, and CH$_3$C$_2$H, to
definitely higher temperature conditions, $\sim$ 100-200 K, obtained from the SiO, SO, and SO$_2$ spectra. 
In other words, at ambient velocities we are probably sampling different layers of the high-density medium
hosting the star forming process, some of them probably heated by the stellar radiation.
On the other hand, for the outflow component we derive two density regimes: about 10$^4$ cm$^{-3}$ from
CS and around 10$^7$ cm$^{-3}$ from SO, whereas the temperatures are 
always quite high: $\simeq$ 100-200 K, indicating regions compressed and heated by shocks.

\section[]{Discussion}

\subsection[]{Abundances}

With the aim of obtaining an estimate of the abundances of the observed species,
the H$_{\rm 2}$ column densities
have been derived from C$^{18}$O for the ambient component and from
CO and C$^{18}$O for the outflow peak, by assuming a standard $X$[CO]/$X$[H$_2$] ratio  
($\simeq$ 10$^{-4}$; Frerking et al. 1982, Lacy et al. 1994).
We obtained the fractional abundances with respect to H$_{\rm 2}$ as the ratio
between column densities and finally an enhancement factor $f$ for each
molecule has been derived as the ratio between the
abundances in the outflow and in the ambient peaks.
The derived values are reported in the last column of Table 4, where two classes  
can be distinguished.
On the one hand, CS, SO, and SO$_2$ do not show a definite abundance enhancement
($f$ $\simeq$ 1), while the HCS$^+$ abundance increases only by
a factor of 4. 
On the other hand, CH$_3$OH, HDO, H$_2$CS, and OCS appear to be definitely enhanced by
at least one order of magnitude ($f$ $\sim$ 10-20),  
confirming that these molecules are closely associated with the chemistry
occurring in molecular outflows (Bachiller \& Per\'ez Guti\'errez 1999). 

Given the well known observational difficulties of detecting the H$_2$O isotopomer,
HDO represents with H$_2^{18}$O an alternative tool to investigate the
occurrence of water emission in space.
In fact, HDO has been already observed towards high-mass star forming regions, mainly 
hot cores (e.g. Jacq et al. 1990, Helmich et al. 1996, Gensheimer et al. 1996 and references therein).
In addition, the high velocity peak indicates that HDO, and thus water, is a very sensitive
shock tracer in molecular outflows driven by YSOs.
The present data do not allow us to give a direct estimate of the H$_2$O abundance.
However, by assuming the cosmic [D]/[H] ratio, $\sim$ 1.5 10$^{-5}$ 
(Oliveira et al. 2003), or the typical
[D]/[H] ratios derived for hot cores, 2-6 10$^{-4}$ (Jacq et al. 1990), 
it is possible to roughly calculate a H$_2$O abundance from our HDO measurements.
We therefore derive for CepA-East $X$[H$_2$O]/$X$[H$_2$] in the 10$^{-6}$-10$^{-4}$ range for
the ambient peak and $\simeq$ 10$^{-5}$-10$^{-4}$ for the outflow peak,  
confirming that the water abundance can be extremely enhanced 
in star forming regions and that H$_2$O is a major coolant in warm gas
(van Dishoeck \& Blake 1998 and references therein).

Finally, the high abundances of CH$_3$OH, HDO, H$_2$CS, and OCS,
which show a well separated high velocity range (see Figs. 10 and 12),
are useful tools to investigate the shocked material by comparing to those
of other shock tracers such as SiO, SO, and SO$_2$.

\subsection[]{Different tracers at different velocities}

In order to investigate the properties of the line wings which
do not show the peak at --5.5 km s$^{-1}$, Fig. 14 reports the distribution with
velocity of the brightness temperature ratio between SO(7--6) (upper panel),
SO$_2$(5$_{24}$--4$_{13}$) (middle panel), H$_2$S(1$_{10}$--1$_{01}$) (lower panel) and
SiO(5--4), as observed towards CepA-East. 
Note that for such a comparison 
we selected transitions observed with a similar HPBW (SO: 9$\arcsec$, SO$_2$: 10$\arcsec$,
H$_2$S and SiO: 11$\arcsec$) and with similar excitation ($E_{\rm u}$ $\sim$ 21-35 K, see Table 1). 
From Fig. 14 it is possible to see that:

\begin{enumerate}

\item
SiO emission dominates at the largest velocities, where 
the highest excitation conditions are expected. Actually, a comparison
between the line profiles of different excitation transitions of
H$_2$S and SO$_2$ observed towards CepA-East clearly indicates
that the higher the velocity the higher the excitation (see Fig. 8 of
Codella et al. 2003);

\item
among the S-bearing species, SO$_2$ has the distribution associated
with the largest width, followed by SO, and finally by H$_2$S.
This is confirmed also by the SO/H$_2$S, SO$_2$/H$_2$S, and
SO$_2$/SO column density ratio profiles, not shown here, which increase 
with velocity.

\end{enumerate}

These results confirm the close association of SiO with the shocks occurring 
along the molecular outflows. Sputtering on dust grains is probably the most
efficient mechanism that leads to a SiO abundance enhancement, although also
grain-grain collisions can play a role depending
on the gas density (Schilke et al. 1997, Caselli et al. 1997).
However, it seems that H$_2$S, SO, and SO$_2$ preferentially
trace more quiescent regions than SiO. from the  
different widths of the resultant column density ratio profiles, 
we note the lack of an intense H$_2$S emission at the highest velocities.
A possible explanation is a fast conversion of H$_2$S into SO and
SO$_2$ and possibly OCS, 
in agreement with the chemical models where first H$_2$S is injected from grains,
then the other S-bearing species are formed very quickly in $\sim$10$^3$ yr
(Pineau des For$\hat {\rm e}$ts et al. 1993, Charnley 1997, Viti et al. 2004).
Alternatively, it may be
that hydrogen sulphide is not the major sulphur carrier in
the grain mantles, as suggested also by the lack of H$_2$S features in the ISO spectra
(Gibb et al. 2000, Boogert et al. 2000) and by the low  
temperature derived from the present H$_2$S spectra, $\sim$ 27 K,
well in agreement with that measured by van der Tak et al. (2003, 25 K) towards
high-mass star forming regions. 

On the other hand, the occurrence of the outflow peak at --5.5 km s$^{-1}$ shows that:

\begin{enumerate}

\item
OCS and H$_2$CS emit at high velocities, where SiO emission starts to dominate and
the excitation conditions are high.
At these velocities also the abundance of CH$_3$OH, which is expected to be abundant in grain mantles
and to be released in the gas phase after shocks, definitely increases. 
These results suggest that OCS and H$_2$CS are among the sulphur carriers
in the grain mantles, or that at least they are rapidly formed once the mantle is evaporated.
Both scenarios are in agreement with the results by van der Tak et al. \shortcite{vander}  
who measured high excitation temperatures for OCS ($\sim$ 100 K). 
Unfortunately, the present data do not allow us to derive a direct temperature estimate from
the OCS data. However, an estimate of the temperature of the gas traced by OCS
is given by the CH$_3$OH emission, which well defines the OCS outflow peak 
and leads to a high temperature: 220 K;

\item
HDO shows an enhancement of its fractional abundance ($X$[HDO]/$X$[H$_2$]) at such high velocities. 
From the estimated value, 10$^{-8}$, we can roughly infer the abundance of water assuming a D/H ratio
equal to 1.5 10$^{-5}$ (Oliveira et al. 2003): $X$[H$_2$O]/$X$[H$_2$] $\simeq$ 6 10$^{-4}$. This
suggests that the majority of oxygen is here locked into water molecules (Meyer et al. 1998).
This is expected if
we are observing regions heated by shocks above 200 K, so that the majority of OH molecules
quickly produce water through reactions with molecular hydrogen  
(e.g. Hartquist et al. 1980). The methanol measurements confirm 
the occurrence of high temperatures at these velocities.

\end{enumerate}

Finally, note that sulphur could be also 
in atomic form when evaporated
from the grain mantles, as suggested by Wakelam et al. \shortcite{wake}. 
However, the atomic S is expected to be very quickly ($\le$ 10$^3$ yr)
locked into SO and SO$_2$, so that a big amount of atomic S is not 
expected to last in the gas phase for a long time and thus difficult to observe.
The form that sulphur takes on dust grains 
is still far to be known: the present data indicate that OCS and H$_2$CS may play
an important role and/or that they are effectively formed in high-temperature gas.

In conclusion, the present analysis of the line profiles indicates that
H$_2$S, SO, and SO$_2$  
may not be easily used as chemical clocks of the shocked material. 
On the other hand, H$_2$CS and OCS could be useful candidates:  
in the next section, a possible use of their emission is discussed. 

\begin{figure*}
\psfig{file=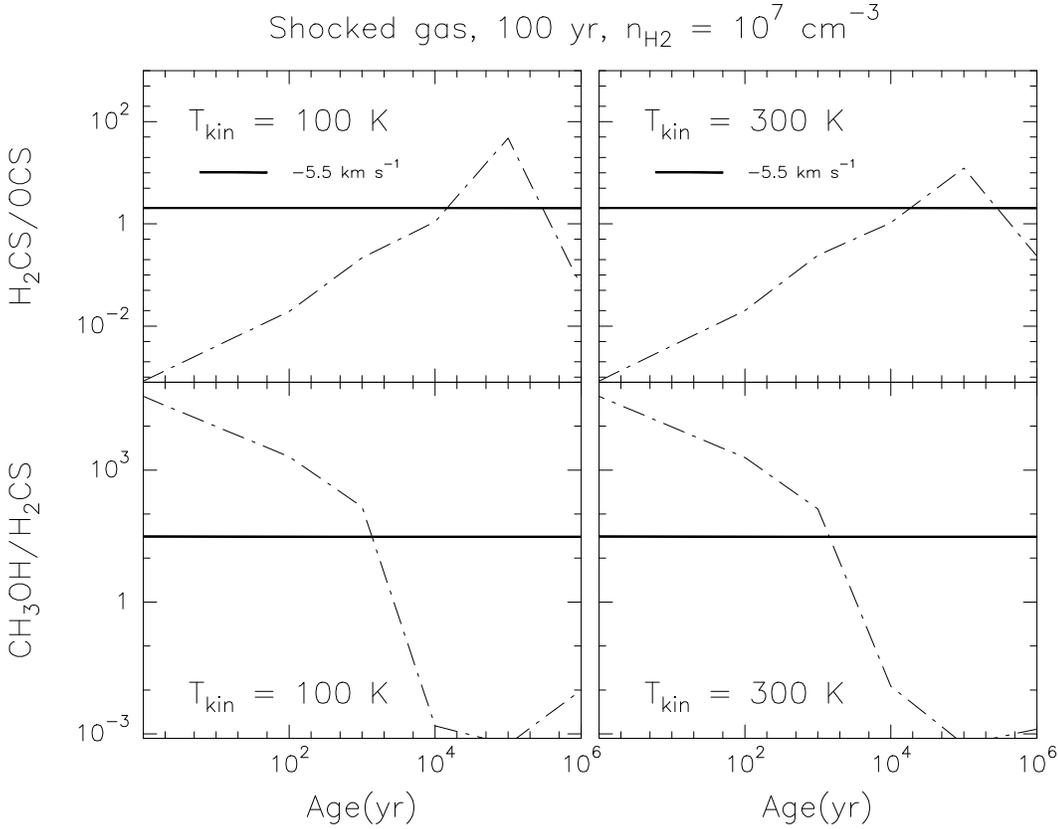,width=14cm,angle=-90}
\caption{Evolution of the abundance ratios H$_2$CS/OCS (upper panels)
and CH$_3$OH/H$_2$CS (lower panels) as functions of time for
a gas shocked for 100 yr (see text), a gas density of
10$^7$ cm$^{-3}$, and gas temperatures
of 100 (left panels) and 300 K (right panels), according 
to the Wakelam et al. (2004) model.
The horizontal thick lines indicate the values derived from the observations
towards CepA-East for the --5.5 km s$^{-1}$ peak
(2 for H$_2$CS/OCS and 30 for CH$_3$OH/H$_2$CS, see Table 4).}
\end{figure*}

\subsection[]{On the origin of the outflow spectral peak}

Since the outflow peak is clearly defined by two sulphuretted species like OCS and H$_2$CS
and by two shock tracers like CH$_3$OH and HDO, we attempt to use
such emission as a chemical clock.
We compared the observations 
with the theoretical calculations recently reported by Wakelam et al. \shortcite{wake}.
The authors developed a time-dependent chemical model 
of sulphur chemistry in hot-cores with up-to-date reaction rate coefficients, following
the molecular composition after the injection of grain mantle species into
the gas phase. They found that the abundances of the main S-bearing species
(H$_2$S, SO, OCS, and SO$_2$) strongly
depend on the physical conditions, on the oxygen abundance in the
gas phase, on the adopted grain mantle composition as well as 
on the time. Hence, the use of the abundance ratios can be often useless to derive age estimates.
Nevertheless, Wakelam et al. \shortcite{wake} compared the observed and predicted abundance
ratios for the hot cores Orion KL and IRAS16293-2422, for which the physical 
conditions had been previously obtained. The authors were able to reproduce the
observed abundance and give an age estimate only assuming that a 
large amount of atomic sulphur is initially
present in the post-evaporative gas, with 
$X$[S]/$X$[H$_2$] between 3 10$^{-5}$ and 3 10$^{-6}$.

In view of these results, we applied the Wakelam et al. \shortcite{wake} model
to the observed emission of the CepA outflow peak. 
Although the model has been
initially used to investigate hot cores, it follows how the S-bearing molecular
abundances vary with time when the gas undergoes a sudden change in its temperature
and density, and in its overall chemical abundance, because of the evaporation of grain
mantles. Therefore, in principle, the model can be reasonably used also for shocked conditions
occurring along a molecular outflow.
However, the physical conditions and the evolution with time should
be different from the hot core case since the physical processes are
different. Magneto-Hydrodynamics models predict that during the passage of a shock wave,
the gas temperature and density simultaneously increase
before decreasing after the shock passage on time scales which depend on
the considered model (e.g. Bergin et al. 1998, Flower \& Pineau des For$\hat {\rm e}$ts 2003).
In our cases, the physical conditions are difficult to constrain in the shocked regions because we
do not know how the physical conditions evolved and also because the
observed molecules indicates temperature and density ranges (see Table 4).
For these reasons, we considered three different models.  
In the first two models, the temperature and the density suddenly increase
to 1000 K and 10$^7$ cm$^{-3}$ and the chemistry evolves for 100 (Case 1) and
1000 yr (Case 2), respectively. After the passage of the shock, the temperature
decreases to the observed values, i.e. $\simeq$ 200 K. The chemistry of the
post-shocked gas then evolves for 10$^6$ yr.
In the third model (Case 3), we assumed that the observed molecules mostly
trace the outer material of the shock where the temperature  and the
density only increase to the observed values 
without passing through a phase with temperatures around 1000 K.
This last scenario is
motivated by the fact that at the CepA distance (725 pc), the filling factor relative to the 
shocked regions where the temperature is expected to exceed 1000 K could
be definitely small and thus their emission could be strongly diluted.

From a chemical point of view, the starting point is the dark molecular
cloud composition computed by the chemical model (see details of
composition A in Wakelam et al. 2004). Due to the shock passage, the
molecules contained on the grain mantles, such as H$_2$O, H$_2$CO,
CH$_3$OH and S-bearing species, are sputtered in the gas phase. 
For the other species, we took the abundances observed as icy features 
towards massive star forming regions (see Wakelam et al. 2004). 

Since the initial form and abundance of sulphur
sputtered from grains is still an open question, as a first step, 
we decided to use the results of Wakelam et al. (2004) who 
reproduce reasonably well the S-bearing observations towards hot cores using a model in which
sulphur is mainly evaporated from grains in atomic, OCS, and H$_2$S forms.
The initial fractional abundances are 10$^{-7}$ for OCS and H$_2$S, 
3 10$^{-6}$ for the S form 
(Palumbo et al. 1997, van Dishoeck \& Blake 1998, Wakelam et al. 2004).
The implicit assumption here is that the majority of sulphur is
depleted in the refractory grain cores.

Since HDO is not included in the Wakelam et al. \shortcite{wake} model, we only used the abundance
ratios between CH$_3$OH, OCS, and H$_2$CS. 
Figure 15 reports the evolution of the H$_2$CS/OCS (upper panels) and
CH$_3$OH/H$_2$CS (lower panels) ratios as a function of time for the
Case 1 (shocked gas 100 yr old). 
The initial abundances of the three species are 9.8 10$^{-8}$, 8.20 10$^{-11}$, and
3.9 10$^{-6}$, for OCS, H$_2$CS, and CH$_3$OH, respectively.
Following the physical parameters
derived for the outflow peak (see Table 4) and adopting a conservative
approach, we used temperatures 
of 100 K (left panels) and 300 K (right panels), whereas for the hydrogen density we assumed
10$^{7}$ cm$^{-3}$ (dot-dashed line). 
The horizontal thick lines indicate the values derived from the observations
towards CepA-East for the --5.5 km s$^{-1}$ peak: 2 for H$_2$CS/OCS and 30 for CH$_3$OH/H$_2$CS. 
Figure 15 shows that, for the assumed $n_{\rm H_2}$ and $T_{\rm kin}$ 
ranges, the observed ratios lead to  
ages for the shocked gas in the CepA-East redshifted component
in the range $\sim$ 10$^{3}$-3 10$^{4}$ yr.  
The H$_2$CS/OCS trend is due to a maximum of the H$_2$CS abundance at
10$^{4}$ yr and to a decrease of the OCS abundance after 10$^{3}$ yr. 
On the other hand, the CH$_3$OH/H$_2$CS trend is due to a dramatic decrease of the
methanol abundance for ages $\ge$ 10$^{4}$ yr.
Note that the CH$_3$OH/H$_2$CS result solves the dichotomy given 
by the H$_2$CS/OCS ratio, which is less selective, 
proposing ages either $\le$ 10$^{4}$ yr or $\sim$ 10$^{6}$ yr. 

Cases 2 and 3 (shocked gas 1000 yr and warmed dark cloud)
do not show significant changes and thus lead to the same conclusion,
leaving open the question about the origin of the outflow peak: gas
which passed through a hot phase or warm gas in the shock surroundings?
In any case, even taking into account all 
the uncertainties of the column density estimates, of the 
abundance ratios, and of the physical conditions,
the comparison between the observations and the
model used here allow us to obtain an age for the shocked gas in CepA-East
of the order of $\sim$ 10$^{3}$-3 10$^{4}$ yr.
Only precise measurements
of density and temperature, which are usually hampered by the high degree
of confusion associated with the observed regions, can refine 
age estimates further.
Future investigations of emission due to S-bearing species towards
other molecular outflows and hot cores where the physical 
conditions are known will confirm whether the CH$_3$OH/H$_2$CS and H$_2$CS/OCS abundance
ratios can in fact be used to estimate ages.

\section[]{Summary}

The CepA star forming region has been investigated through a multiline
survey at mm-wavelengths. 
The results indicate the occurrence of a rich chemistry surrounding
the YSOs of the CepA-East stellar association.
The main findings are as follows: 

\begin{enumerate}

\item[1.] The CS and HDO maps draw a complex scenario, detecting high excitation clumps hosting 
the YSOs driving multiple outflows. Four main flows have been identified:  
three are along the SW, NE, and SE directions, traced also by strings of VLA continuum sources,
and are accelerating high density CS clumps. 
In addition, HDO reveals a fourth outflow pointing towards South,
which had been previously detected only through H$_2$S and SO$_2$ observations,
and is then associated with conditions particularly favourable to a
chemical enrichment.

\item[2.] At the CepA-East position 
different molecules exhibit different spectral behaviours: 
three classes can be identified. 
Some species (C$^{13}$CH, C$_3$H$_2$, CH$_2$CO, CH$_3$C$_2$H, HC$^{18}$O$^+$)
peak with relatively narrow lines at ambient velocities (ambient peak). 
Other molecules (CO, CS, H$_2$S, SiO, SO, SO$_2$)
show extended wings and trace the whole range of the outflow velocities. Moreover, 
there is a group of species (OCS, H$_2$CS, HDO, and CH$_3$OH) which 
shows wings and, in addition, well defines a high
velocity redshifted spectral peak (outflow peak) 
which can be used to investigate the SE-S outflows. 

\item[3.]The physical conditions associated with both the ambient and outflow peaks have been estimated.
By means of statistical-equilibrium calculations, using LVG codes, we have
obtained the physical
parameters of the molecular gas traced by SO and CS. For 
CH$_3$C$_2$H, H$_2$CS, $^{34}$SO$_2$, and CH$_3$OH we have used 
the rotation diagram method.
For the other molecular species, we have 
calculated the total column densities assuming LTE conditions.
At ambient velocities the gas is quite dense ($>$ 10$^5$ cm$^{-3}$)
and different components at different temperatures coexist, ranging from the
relatively low kinetic temperatures ($\le$ 50 K)
measured with H$_2$S, CH$_3$OH, H$_2$CS, and CH$_3$C$_2$H, to
definitely higher temperature conditions, $\sim$ 100-200 K, 
obtained from the SiO, SO, and SO$_2$ spectra, which may trace layers directly
heated by the stellar radiation. 
For the outflow component densities between $\sim$ 10$^4$ cm$^{-3}$ to
$\sim$ 10$^7$ cm$^{-3}$ and high temperatures, $\simeq$ 100-200 K, have
been found, indicating regions compressed and heated by shocks.

\item[4.]The comparison between the line profiles of different outflow tracers of molecular
outflows shows that SiO dominates at the highest velocities, where the highest
excitation conditions are found. This confirms the close association of SiO with shocks.
On the other hand, H$_2$S, SO$_2$, and SO preferentially trace more quiescent regions.
In particular, we find a lack of a bright H$_2$S emission at the highest velocities.
Moreover, OCS and H$_2$CS emit at quite high velocities, where (i) SiO emission dominates
and the excitation conditions are high, and (ii) CH$_3$OH and HDO, other shock tracers, increase 
their abundance.
These results could indicate that H$_2$S is not the only major sulphur carrier in the grain
mantles,
and that OCS and H$_2$CS may probably play an important role on the grains; or that 
alternatively they could rapidly form once the mantle is evaporated after the passage of a shock.

\item[5.]We checked the possible use of the CH$_3$OH, OCS, and H$_2$CS emission 
as chemical clocks to measure the age of the shocked material
associated with the outflow peak.
The observations have been compared with the theoretical calculations recently 
reported by Wakelam et al. \shortcite{wake}, who developed a time-dependent model
with up-to-date reaction rate coefficients for the sulphur chemistry.
Once associated with the derived physical parameters, 
the H$_2$CS/OCS and CH$_3$OH/H$_2$CS column density ratios led to
ages in the range 10$^{3}$-3 10$^{4}$ yr. 
Further observations of emission due to sulphuretted
molecules as well as accurate measurements of density and temperatures associated
with the outflowing material are necessary in order to calibrate these chemical clocks.

\end{enumerate}

\section*{Acknowledgements}

We are grateful to J. Santiago Garcia, M. Tafalla, C. M. Walmsley, and D. A. Williams 
for helpful suggestions and discussions. SV acknowledges individual financial support
from a PPARC Advanced Fellowship. We also thank the referee for the contribution to the
improvement of the paper.

\end{document}